# Methods for Jet Studies with Three-Particle Correlations

Claude A. Pruneau
Physics and Astronomy Department, Wayne State University
666 West Hancock, Detroit, MI 48152 USA

We present a method based on three-particle azimuthal correlation cumulants for studying jet interactions with the medium produced in heavy ion collisions (at RHIC) where jets cannot be reconstructed on an event-by-event basis with conventional jet-finding algorithms. The method is specifically designed to distinguish a range of jet interaction mechanisms such as Mach cone emission, gluon Cerenkov emission, jet scattering, and jet broadening. We describe how anisotropic flow contributions of second order (e.g. $v_2^2$) are suppressed in three-particle azimuthal correlation cumulants, and discuss specific model representations of di-jets, away-side scattering, and Mach cone emission.

## 1. Introduction

Recent measurements at RHIC unraveled the production of strongly interacting quark gluon plasma (sQGP) in Au + Au collisions. This conclusion is based on the observation in Au + Au collisions of large collective flow, in agreement with hydrodynamic calculations, and dramatic suppression of particle production at high transverse momentum relative to expectations based on p + p interactions scaled by the number of binary collisions in Au + Au[1,2,3]. Comparison of azimuthal two-particle correlations measured in Au+Au, d+Au, and p+p collisions indicates the production of jets is strongly modified by their propagation through dense matter produced in Au + Au. Modifications are manifested for low $p_t$ jets (e.g. $3 < p_t < 4$ GeV/c) by a complete disappearance of the away-side jet [4,5], and for higher pt jets by a large suppression of the associated particle production on the away-side jet. Various mechanisms are proposed in the recent literature to explain the observed jet modifications. Typically, these assume the trigger jet is produced near the surface of the dense matter (produced by the colliding nuclei) and thus escape mostly unscathed while the away-side jet has to penetrate through the dense matter and can therefore undergo a number of interactions. Attenuation and modification mechanisms range from multiple scattering of the initial parton, gluon radiation before and after fragmentation [6,7,8,9]. Recently, the PHENIX collaboration reported it observed a dip and anomalous peak structures in two-particle correlations for trigger particle in the range 2.5-4 GeV/c and associates in the range 1 - 2.5 GeV/c [10]. A number of authors argue conical flow (also called Mach cone emission) caused by the away-side jet [11,12] or Cerenkov effect (gluon radiation) may be responsible for the dip and peak structures [13]. The issue is, however, strongly debated. Authors of Ref 14 argue the jet energy loss is too small and jet interaction with the medium should lead to very small effects, while Voloshin[15] argues the observed correlation can be due to jet flow caused by the underlying flowing matter.

The interpretation of two-particle correlations in A+A collisions is complicated by the fact their shapes are rather sensitive to the measured $p_t$ ranges. The STAR collaboration reported at QM05 the progressive re-appearance of away-side jets when higher pt trigger particles are selected (i.e., for $p_t > 8$ GeV). While the changing shape of the two particle correlations may arise, in part, because of the $p_t$ dependence of jet interaction with the medium we seek to understand, it may also result from the interplay of various, less interesting (in this specific context) particle production processes such as resonance decays, radial and elliptical flow, momentum, and quantum numbers conservation. It is thus desirable to reduce ambiguities by performing a more complete set of measurements. Since explicit reconstruction of jets, event-by-event, is impractical in Au + Au collisions, one is limited to correlations studies. We contend that one can gain additional insight into the jet interaction with the medium through measurements of three-particle correlations.

Measurements of three-particle correlations should enable a straightforward elimination of a simple two-particle process and enable unambiguous identification of genuine multi-particle production phenomena e.g., jets, Mach cone, etc. Measurements of three-particle correlations should also, in principle, enable one to distinguish some of the proposed jet attenuation mechanisms. Measuring a triplet of particles' yield alone is, however, not sufficient to actually eliminate contributions from two-body processes and collective processes (a.k.a. flow). Additional steps must be taken to eliminate such effects from measured triplet cross-sections.

In this paper, we describe a measurement technique based on cumulants and "predict" the measurement outcome for some simple key models of particle production. The technique is described in detail in Section 2. We then apply the technique to simple particle production models that can be computed analytically, or through Monte Carlo calculations in Section 3. Our results are summarized and discussed in Section 4.

## 2. The Cumulant Method

Correlation measurements are based on the simultaneous measurement of two-, three- or n-particles. In such measurements, one has no a-priori knowledge of the process or processes leading to the production of these particles. Indeed one may be dealing with one, two, or many (distinct or not) production processes: radial flow, elliptical flow, resonance decay, jets, Mach cone, etc. Given these processes are intrinsically stochastic (random) in nature, it is not possible a-priori, to determine which of such processes lead to production of a specific pair or triplet of particles. Members of a given triplet of measured particles may be produced by one of many multi (n>2) particle production processes, or a combination of such processes. More explicitly stated, it is possible that all three particles of a triplet are correlated because the same process produced them. It is, however, also possible that only two of the three particles are actually from the same process, while the third is from an independent, uncorrelated process. Additionally, it is also possible that all three particles of given triplet were produced by different and independent processes. For the study of jet related phenomena, we are interested in identifying those triplets that consist of three particles produced by the same process or phenomenon. Unfortunately, given the stochastic nature of particle production in nuclear interactions, it is obviously impossible to distinguish on a pair by pair, or triplet-by-triplet basis, which are truly correlated, i.e. from the same process, than those amounting to random combinations. It is therefore necessary to utilize statistical techniques, only valid for an ensemble of triplets (or pairs), when endeavoring to separate the different processes contributing to particle production. The cumulant method is specifically designed to accomplish this task.

Cumulants were introduced by Berger[16] and discussed by Carruthers *et al.* [17] and are now used in a variety of analyses[18]. We summarize here the definition and essential properties of two- and three-particle cumulants relevant for the discussion in this work. We restrict the discussion and notation to azimuthal angle measurements. We note single particle densities by $\rho_1(\varphi_i) = dN/d\varphi_i$, two particle densities $\rho_2(\varphi_i,\varphi_j) = dN/d\varphi_i d\varphi_j$, and three-particles densities with $\rho_3(\varphi_i,\varphi_j,\varphi_k) = dN/d\varphi_i d\varphi_j d\varphi_k$. Measurements of "n-plet" of particles may yield "n" truly correlated particles, "n-1" truly correlated particles in combination with an uncorrelated one, "n-2" truly correlated in combination with a pair of particles from one or two other processes, etc. In the case of a measurement of two and three-particle densities, this can be written

$$\rho_2(\varphi_i,\varphi_j) = \hat{\rho}_2(\varphi_i,\varphi_j) + \rho_1(\varphi_i)\rho_1(\varphi_j)$$
$$\rho_3(\varphi_i,\varphi_j,\varphi_k) = \hat{\rho}_3(\varphi_i,\varphi_j,\varphi_k) + \hat{\rho}_2(\varphi_i,\varphi_j)\rho_1(\varphi_k) + \hat{\rho}_2(\varphi_i,\varphi_k)\rho_1(\varphi_j) + \hat{\rho}_2(\varphi_j,\varphi_k)\rho_1(\varphi_i)$$
$$+\rho_1(\varphi_i)\rho_1(\varphi_j)\rho_1(\varphi_k) \quad (2.1)$$

The truly correlated particle densities, indicated with "^", are obtained by solving the two equations above for $\hat{\rho}_2(\varphi_i,\varphi_j)$ and $\hat{\rho}_3(\varphi_i,\varphi_j,\varphi_k)$. One obtains the definition of the 2- and 3-cumulants in terms of measured densities.

$$\hat{\rho}_2(\varphi_i,\varphi_j) = \rho_2(\varphi_i,\varphi_j) - \rho_1(\varphi_i)\rho_1(\varphi_j)$$
$$\hat{\rho}_3(\varphi_i,\varphi_j,\varphi_k) = \rho_3(\varphi_i,\varphi_j,\varphi_k) - \rho_2(\varphi_i,\varphi_j)\rho_1(\varphi_k) - \rho_2(\varphi_i,\varphi_k)\rho_1(\varphi_j) - \rho_2(\varphi_j,\varphi_k)\rho_1(\varphi_i)$$
$$+2\rho_1(\varphi_i)\rho_1(\varphi_j)\rho_1(\varphi_k) \quad (2.2)$$

In this work, we consider particle production as arising from a superposition of "s" independent processes such as collective flow, two-body decays, jet production, etc. Identifying these processes on the basis on a generic index α, with α = 1, ..., s, it is straightforward to show that if all processes, α, are statistically independent, the single particle density and measured 2- and 3- cumulants may be expressed as a sum of the cumulants of each of these "s" independent processes.

$$\hat{\rho}_1(\varphi_i) = \sum_{\alpha=1}^{s} \hat{\rho}_{1,s}(\varphi_i)$$
$$\hat{\rho}_2(\varphi_i,\varphi_j) = \sum_{\alpha=1}^{s} \hat{\rho}_{2,s}(\varphi_i,\varphi_j) \quad (2.3)$$
$$\hat{\rho}_3(\varphi_i,\varphi_j,\varphi_k) = \sum_{\alpha=1}^{s} \hat{\rho}_{3,s}(\varphi_i,\varphi_j,\varphi_k)$$

where $\hat{\rho}_{3,\alpha}(\varphi_i,\varphi_j,\varphi_k)$ and $\hat{\rho}_{2,\alpha}(\varphi_i,\varphi_j)$ correspond to the two- and three-particle cumulants for process "α". This applies whether one deals with independent processes of the same or different types. If there are, on average, N rho-meson decay per collision, then the cumulant associated with these shall be simply N times the cumulant of one rho decay.

Experimentally, cumulant measurements are subject to the same limitations associated with finite acceptance and detection efficiency as those involved in measurements of single particle densities. While it is not possible to compensate for finite acceptance in a model-independent way, one can express the cumulants in terms of probability densities, rather than densities, and obtain experimentally robust quantities. While it is, in principle, possible to perform detailed calculations of the single, $\varepsilon_1(\varphi_i)$, pair, $\varepsilon_2(\varphi_i,\varphi_j)$ and triplet, $\varepsilon_3(\varphi_i,\varphi_j,\varphi_k)$, efficiencies for the all- phase space of interest (based on Monte Carlo simulations of the detector response) such simulations may become prohibitively CPU expensive in practice. However, to the extent that it is reasonable to assume that the two- and three-particle detection efficiencies can be factorized as the product of single particle efficiencies, e.g. $\varepsilon_3(\varphi_i,\varphi_j,\varphi_k) \approx \varepsilon_1(\varphi_i)\varepsilon_1(\varphi_j)\varepsilon_1(\varphi_k)$, then ratios of two- and three-particle cumulants to products of two and three single particle densities yield robust experimental quantities:

$$\frac{\rho_2^M(\varphi_i,\varphi_j)}{\rho_1^M(\varphi_i)\rho_1^M(\varphi_j)} = \frac{\varepsilon_2(\varphi_i,\varphi_j)}{\varepsilon_1(\varphi_i)\varepsilon_1(\varphi_j)} \frac{\rho_2^A(\varphi_i,\varphi_j)}{\rho_1^A(\varphi_i)\rho_1^A(\varphi_j)} \simeq \frac{\rho_2^A(\varphi_i,\varphi_j)}{\rho_1^A(\varphi_i)\rho_1^A(\varphi_j)}$$
$$\frac{\rho_3^M(\varphi_i,\varphi_j,\varphi_k)}{\rho_1^M(\varphi_i)\rho_1^M(\varphi_j)\rho_1^M(\varphi_k)} = \frac{\varepsilon_3(\varphi_i,\varphi_j,\varphi_k)}{\varepsilon_1(\varphi_i)\varepsilon_1(\varphi_j)\varepsilon_1(\varphi_k)} \frac{\rho_3^A(\varphi_i,\varphi_j,\varphi_k)}{\rho_1^A(\varphi_i)\rho_1^A(\varphi_j)\rho_1^A(\varphi_k)} \quad (2.4)$$
$$\simeq \frac{\rho_3^A(\varphi_i,\varphi_j,\varphi_k)}{\rho_1^A(\varphi_i)\rho_1^A(\varphi_j)\rho_1^A(\varphi_k)}$$

where "M" and "A" denote measured and actual quantities respectively. An alternative approach to account for finite detection efficiencies is the use of the "mixed events" technique first introduced by Kopylov[19]. Unfortunately, with that technique, the absolute normalization of the correlations may be lost. There are also issues of reliability connected to the necessity of preserving the event multiplicity, net charge, and other conserved quantity distributions.

The recent discovery of disappearance and reappearance[20] of jets reported at RHIC were based on two-particle correlation studies. Much was, and can still, be learned from two-particle correlations. The interpretation of two-particle correlations is, however, somewhat ambiguous: structures found in analyses reported by PHENIX[10], and STAR may be interpreted as resulting from parton scattering, Mach cone emission[11, 12, 14, 27], Cerenkov gluon radiation[28], jet flow[15], and possibly other mechanisms as well. We argue that ambiguities can be reduced, and jet properties further studied with three-particle correlations. Consider, for instance, the observation of away-side jet broadening. With two particle correlations, the high pt "trigger" or "tag" usually defines the direction of the near-side jet. The second particle is generally assumed to come from the away–side jet or from some associated process (e.g. Mach cone emission). Observations with low $p_t$ particles have shown that the away-side is typically much broader than the near-side or shows a dip at 180º and exhibits side peaks near 120º and 240º. Observations also reveal the width of the away-side peak decreases for increasing $p_t$. The problem then arises that it is not possible, based on a two-particle correlation, to distinguish whether the broadening of the away–side is due to scattering of the jet leading parton with no actual broadening of the jet, as schematically illustrated in Figure 1(c), or due to the interactions and dispersion of jet fragments by the medium shown in Figure 1 (b). The ambiguity is eliminated in three-particle correlations by measuring the width of the correlation between two away-side particles. If the jet structure is unchanged except for initial scattering of the leading jet parton, then the width of the away-side particles remains unchanged between p + p and Au + Au collisions. If, on the other hand, the jet fragments are dispersed (scattered) by the medium, then the width of the away-side should indeed increase as schematically illustrated in Figure 1 (b). These two scenarios are discussed on the basis of a toy model in Sections 3.2 and 3.5. Three-particle azimuthal correlations shall also be useful for identifying Mach cone emission or Cerenkov radiation. In the case of predicted Mach cone emission, the propagation of the away-side parton in the dense medium formed by the A+A collisions leads to the production of a wake at an angle determined by the ratio of parton speed and sound velocity in the medium. For a QGP plasma, the sound velocity is expected[25] to be on the order of 0.33c whereas the parton speed is near c. This may then lead to particle emission at 60-70º from the away-side direction[25], as illustrated in Figure 1 (d,e). While the number of particles emitted in the Mach cone is expected to be fairly modest, it should be nonetheless possible, with appropriate kinematical cuts, to identify Mach cone emission in three-particle correlations by the emergence of four-side structures 60-70º from the away-side direction as schematically illustrated in Figure 1 (See however the discussion in Sect. 3.8). By contrast, parton deflection (scattering) should instead lead to an elongation along the diagonal of the away-side jet peak. Cerenkov gluon emission is predicted to have similar three-particle emission structures as Mach cone emission, although the characteristics of this emission are not as clearly understood. Three-particle correlations might also be useful to identify the production of a backsplash by the away-side parton. Authors of reference 14 have argued, based on transport calculations conducted with a 2D hydro calculation, that due to radial outward flow, and given the relatively small energy deposited by the away-side parton, the production of a Mach cone is rather unlikely. They instead discuss the possibility of a backsplash. Such a backsplash might be visible as an excess of particle production at small angles relative to the near side peak.

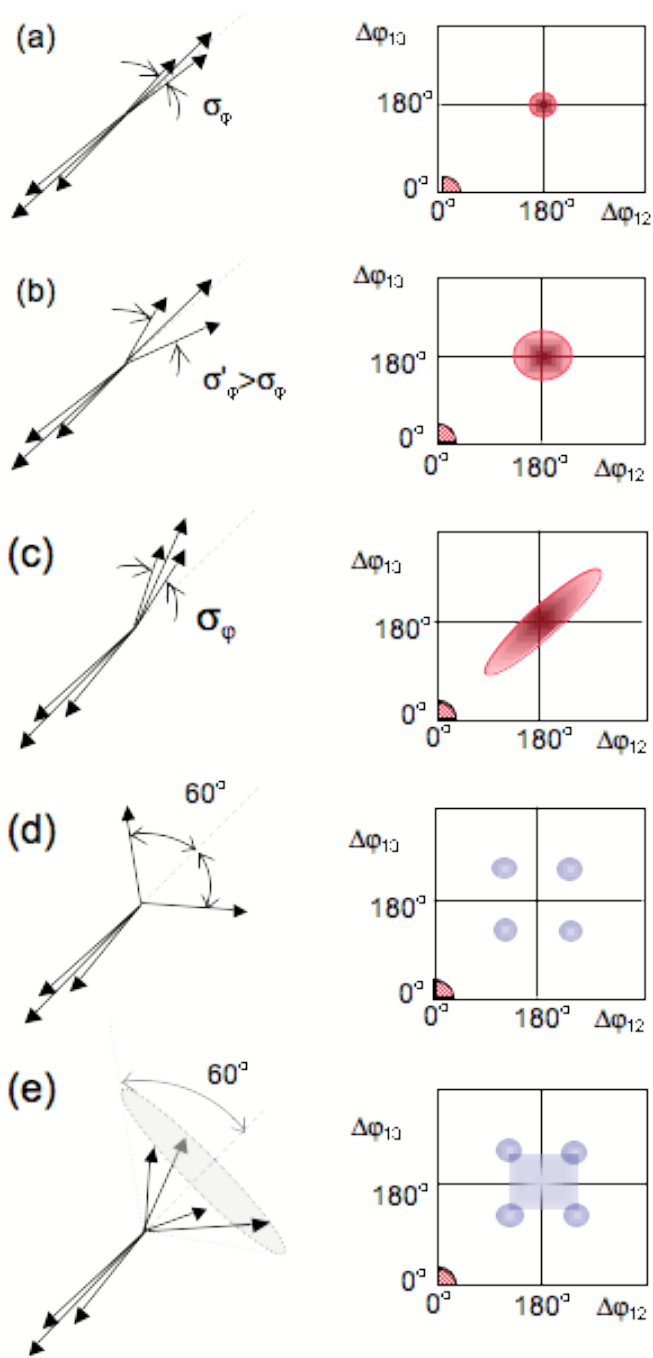

Figure 1 Left: Illustration in transverse plane of (a) in vacuum back-to-back di-jet event, (b) in bulk broaden away-side di-jet event, (c) deflected jet, (d) Mach cone with particle emission perpendicular to the beam direction, and (e) Mach cone with particle emission at all azimuths relative to the away-side parton direction. Right: Schematic illustration of the three-particle correlations expected for each mechanism shown on the left. See text for details. (in color online).

We thus discuss the formulation of three-particle cumulants as correlations based on two azimuthal angle differences. Here we will illustrate the method with differences between azimuthal angles of particle 1 and 2, $\Delta\varphi_{12}$, and particle 1 and 3, $\Delta\varphi_{13}$. The technique is trivially extended to full phase space. The remainder of this section presents a discussion of experimental techniques used in the determination of cumulants. Specific models relevant for the study of away-side parton scattering, Mach cone emission, and backgrounds are discussed in Section 3.

Calculating the 3-cumulant and the normalization to singles, in terms of these two variables, requires the calculation of terms such as $\rho_2\rho_1(\Delta\varphi_{12}, \Delta\varphi_{13})$, and $\rho_1\rho_1\rho_1 (\Delta\varphi_{12}, \Delta\varphi_{13})$. First consider the calculation of the $\rho_1\rho_1\rho_1 (\Delta\varphi_{12}, \Delta\varphi_{13})$ term. Calculating this term requires the knowledge of the singles density $\rho_1(\varphi_1)$, $\rho_1(\varphi_2)$, and $\rho_1(\varphi_3)$ with $\varphi_1$, $\varphi_2$, and $\varphi_3$ being the azimuthal angle at which each of the particles are measured. The index 1 refers to the trigger particle, and indices 2 & 3 refer to the associate particles. Kinematic ranges for each of these variables may be chosen arbitrarily to be identical, distinct, or to partially overlap. Here we will assume the trigger range is distinct, while the two associates have identical kinematic ranges. Typically, at a given collision centrality, the single particle distributions may be measured with 1D profile histograms with finitely many bins, n, for azimuthal angles between 0 and 360 degrees. The $\rho_1\rho_1 (\Delta\varphi_{12})$ and $\rho_1\rho_1\rho_1 (\Delta\varphi_{12}, \Delta\varphi_{13})$ terms are then calculated using respectively 1D and 2D histograms or arrays using a binning of n and n x n in $\Delta\varphi_{12}, \Delta\varphi_{13}$ and summing over all combinations $\varphi_1$, $\varphi_2$, and $\varphi_3$ that yield the same angle differences:

$$\rho_1\rho_1(\Delta\varphi_{12}) \equiv \rho_1\rho_1(m) = \sum_{i,j=1}^{n} \rho_1(i)\rho_1(j)\delta(m-i+j)$$

$$\rho_1\rho_1\rho_1(\Delta\varphi_{12}, \Delta\varphi_{13}) \equiv \rho_1\rho_1\rho_1(m,p) = \sum_{i,j,k=1}^{n} \rho_1(i)\rho_1(j)\rho_1(k)\delta(m-i+j)\delta(p-i+k)$$

(2.5)

with m, p = 1, ..., n. Calculating the $\rho_2\rho_1 (\Delta\varphi_{12}, \Delta\varphi_{13})$ terms requires the knowledge of $\rho_2(\varphi_i, \varphi_j)$, and $\rho_1(\varphi_k)$. The term $\rho_2(\varphi_i, \varphi_j)$ may be measured using a 2D profile histogram as for the singles. The difference, of course, is that we need to account for which of the three particles are paired. Again, here one uses n x n binning, but in actual angles rather than angle differences. The three $\rho_2\rho_1 (\Delta\varphi_{12}, \Delta\varphi_{13})$ terms are obtained by folding $\rho_2(\varphi_i, \varphi_j)$, and $\rho_1(\varphi_k)$ using the following formula:

$$\rho_2\rho_1(\Delta\varphi_{12}, \Delta\varphi_{13})_{123} \equiv \rho_2\rho_1(m,p) = \sum_{i,j,k=1}^{n} \rho_2(i_1, j_2)\rho_1(k_3)\delta(m-i_1-j_2)\delta(p-i_1-k_3)$$

$$\rho_2\rho_1(\Delta\varphi_{12}, \Delta\varphi_{13})_{231} \equiv \rho_2\rho_1(m,p) = \sum_{i,j,k=1}^{n} \rho_2(i_2, j_3)\rho_1(k_1)\delta(m-i_1-j_2)\delta(p-i_1-k_3)$$

(2.6)

$$\rho_2\rho_1(\Delta\varphi_{12}, \Delta\varphi_{13})_{132} \equiv \rho_2\rho_1(m,p) = \sum_{i,j,k=1}^{n} \rho_2(i_1, j_2)\rho_1(k_3)\delta(m-i_1-j_2)\delta(p-i_1-k_3)$$

with m, p = 1, ..., n. The sub index "123", "231", and "132" notation is used above to distinguish which of the three particles are taken from the two-particle density $\rho_2(\varphi_i\varphi_j)$.

The three-particle density $\rho_3(\Delta\varphi_{12}, \Delta\varphi_{13})$ may be obtained in a number of ways. The most straightforward technique is to use three nested loops and consider, for each event entry, all possible triplet permutations and fill a 2D profile histogram or array. We note that for efficiency correction purposes (which require division of the cumulant terms by singles), best results are achieved using the same binning and integer arithmetic used in computing the cumulant and singles term $\rho_1\rho_1\rho_1 (\Delta\varphi_{12}, \Delta\varphi_{13})$.

We emphasize that 3-cumulants are indeed a measure of the degree of three-particle correlation between three measured particles. As such, they may be zero, positive or even negative. This is true for both differential values of the cumulant (as in the above expression) and integrals of the cumulant over phase space. We illustrate this point as follows. Consider the numbers of particles, $n_i$, measured in (arbitrary) kinematic bins i=1, 2, and 3 in any given event. Clearly these can be expressed as a sum of the means $\langle n_i \rangle$, obtained by averaging over many events, plus some "small" random quantity $r_i$ which varies event-by-event.

$n_i = \langle n_i \rangle + r_i$,

By definition of the mean $\langle n_i \rangle$, one has $\langle r_i \rangle = 0$. The average number of pairs $\langle n_1 n_2 \rangle$ and triplets $\langle n_1 n_2 n_3 \rangle$ are thus:

$$\langle n_1 n_2 \rangle = \langle (\langle n_1 \rangle + r_1)(\langle n_2 \rangle + r_2) \rangle = \langle n_1 \rangle \langle n_2 \rangle + \langle r_1 r_2 \rangle \tag{2.7}$$

$$\begin{aligned}\langle n_1 n_2 n_3 \rangle &= \langle (\langle n_1 \rangle + r_1)(\langle n_2 \rangle + r_2)(\langle n_3 \rangle + r_3) \rangle \\ &= \langle n_1 \rangle \langle n_2 \rangle \langle n_3 \rangle + \langle n_1 \rangle \langle r_2 r_3 \rangle + \langle n_2 \rangle \langle r_1 r_3 \rangle + \langle n_3 \rangle \langle r_1 r_2 \rangle + \langle r_1 r_2 r_3 \rangle \end{aligned} \tag{2.8}$$

Applying the definition (2.2) with the above, one finds the 3-cumulant reduces to $\langle r_1 r_2 r_3 \rangle$, and indeed verifies it can be positive, negative, or null depending on the nature of the particle production processes involved.

The cumulant calculation may be performed inclusively for a wide range of collision centralities or semi-inclusively. Semi-inclusive measurements are carried at fixed reference multiplicity. Averages over large centrality ranges (or bins) of semi-inclusive measurements may be achieved with a simple or cross-section weighted average across centrality bins. Assuming here the centrality is estimated on the basis of some reference particle multiplicity, m, measured over a given experimental acceptance, one gets:

$$C_3(\Delta\varphi_{12}, \Delta\varphi_{13}) = \hat{\rho}_3(\Delta\varphi_{12}, \Delta\varphi_{13})_{BIN} = \frac{\sum_{m=\min}^{\max} w(m) \hat{\rho}_3(\Delta\varphi_{12}, \Delta\varphi_{13})_m}{\sum_{m=\min}^{\max} w(m)} \tag{2.9}$$

where the sub-indices "BIN" and "m" refer to the bin average, and fixed reference multiplicity respectively. The weights, w(m), can be taken as unity for simple arithmetic average or as the number of events at the given reference multiplicity for cross section weighted averages.

In the following, we will use the notations $C_2(\Delta\varphi_{12})$, and $C_3(\Delta\varphi_{12}, \Delta\varphi_{13})$ to identify the 2- and 3-cumulants average over collisions as defined per Eq. 2.5. Experimentally, this quantity is subject to finite efficiencies. It is thus convenient to consider the ratio of $C_2$ and $C_3$ to products of single particle densities to cancel out efficiencies as discussed above. We use the notation $R_k^s$, $R_k^i$ to respectively identify (semi-)inclusive and inclusive averages over collision centrality of the ratio of k-cumulants and products of single particle densities. $R_3^s$, and $R_3^i$ are defined as follows:

$$\begin{aligned} R_3^s(\Delta\varphi_{12}, \Delta\varphi_{13}) &== \frac{\sum_{m=\min}^{\max} w(m) \frac{\hat{\rho}_3(\Delta\varphi_{12}, \Delta\varphi_{13})_m}{\rho_1 \rho_1 \rho_1 (\Delta\varphi_{12}, \Delta\varphi_{13})_m}}{\sum_{m=\min}^{\max} w(m)} \\ \\ R_3^i(\Delta\varphi_{12}, \Delta\varphi_{13}) &== \frac{\sum_{m=\min}^{\max} w(m) \hat{\rho}_3(\Delta\varphi_{12}, \Delta\varphi_{13})_m}{\sum_{m=\min}^{\max} w(m) \rho_1 \rho_1 \rho_1 (\Delta\varphi_{12}, \Delta\varphi_{13})_m} \end{aligned} \tag{2.10}$$

with similar definitions for $R_2^s$, and $R_2^i$. Given the quantity $\rho_3 / \rho_1 \rho_1 \rho_1$ amounts to a probability density, the semi-inclusive normalized cumulant $R_k^s$ is thus a cumulant of probabilities densities. While the interpretation of $R_k^i$ is not as straightforward, it is easier to calculate given finite statistics. We note that because detection efficiencies are, in general, a function of the detector occupancy (and hence particle multiplicity) finite efficiencies involved in measuring $\rho_3$ and $\rho_1\rho_1\rho_1$ do not, in general, cancel in the expression of $R_k^i$ but do for $R_k^s$. An efficiency-corrected particle density cumulant is obtained by multiplying the ratio $R_2^e$ and $R_3^e$ by the angle averaged products $\overline{\rho_1\rho_1}$ and $\overline{\rho_1\rho_1\rho_1}$ respectively. The single particle densities $\rho_1(\varphi)$ can be efficiency corrected via standard techniques such as event embedding. It is thus straightforward to compute the angle averaged $\overline{\rho_1\rho_1}$ and $\overline{\rho_1\rho_1\rho_1}$ and their products to $R_2^e$ and $R_3^e$ respectively, thereby yielding efficiency-corrected particle density cumulants. Again, in this case, the calculation may be performed inclusively or semi-inclusively. We use the notations $S_k^i$ and $S_k^s$ for the inclusive and semi-inclusive cumulants respectively. These are defined as follows:

$$S_3^s(\Delta\varphi_{12},\Delta\varphi_{13}) = \frac{\sum_{m=\min}^{\max} w(m)\overline{\rho_1\rho_1\rho_1}(m)\frac{\hat{\rho}_3(\Delta\varphi_{12},\Delta\varphi_{13})_m}{\rho_1\rho_1\rho_1(\Delta\varphi_{12},\Delta\varphi_{13})_m}}{\sum_{m=\min}^{\max} w(m)}$$

$$S_3^i(\Delta\varphi_{12},\Delta\varphi_{13}) = \frac{\sum_{m=\min}^{\max} w(m)\overline{\rho_1\rho_1\rho_1}\; \sum_{m=\min}^{\max} w(m)\hat{\rho}_3(\Delta\varphi_{12},\Delta\varphi_{13})_m}{\sum_{m=\min}^{\max} w(m)\; \sum_{m=\min}^{\max} w(m)\rho_1\rho_1\rho_1(\Delta\varphi_{12},\Delta\varphi_{13})_m}$$

(2.11)

We note once again that while the calculation of $S_3^i(\Delta\varphi_{12},\Delta\varphi_{13})$ is simpler and requires less storage than the calculation of the semi-inclusive quantity $S_3^s(\Delta\varphi_{12},\Delta\varphi_{13})$, the efficiency correction on $S_3^i(\Delta\varphi_{12},\Delta\varphi_{13})$ is less accurate given the efficiency is typically a function of the event multiplicity.

### 3. Multi-particle Correlations Toy Models

The structure of three-particle correlations can be rather complicated. It is thus useful to consider a simple analytical model to guide one's intuition in the study of jet properties in A + A collisions, based on two- and three- particle correlation studies. We consider, for illustrative purposes, a range of simple models. We begin in section 3.1, with pencil-like correlations and consider collisions producing a superposition of pencil-like jets of particles, i.e., jets consisting of particles emitted in the same direction. We proceed, in section 3.2, to consider a semi-realistic case of Gaussian- shaped jets of particles mixed with background particles with flat distribution. We discuss the important case of harmonic anisotropies in section 3.3. In section 3.4, we consider the production of jets simultaneous to harmonic flow. We use the result of these sections to study di-jet, Mach cone production and similar phenomena in 3.5. These results are contrasted to simulations of two-body (ρ-meson) decay in Section 3.6, where we illustrate through simple examples, that 2-particle correlations may have a wide-range of shapes, depending on the kinematic cuts used to construct them. We also show how the complicated, 3-particle density obtained with ρ-meson decays are reduced to a null signal in 3-particle azimuthal correlation cumulant. Seemingly independent processes, such as anisotropic flow and jet-like Gaussian structures found in two-particle correlations, may have a common origin. We describe in Section 3.7 how differential attenuation of jets by the medium may produce both of these types of correlation structures simultaneously. Finally, we discuss the case of Mach cone emission in Section 3.8.

#### 3.1. Pencil-Like Jets (Model PJ)

We begin with a simple, albeit unrealistic, jet model where we assume all jet fragments are emitted exactly in the direction of the parton initiating the jet. The direction of the jet (parton) is chosen to have a flat probability distribution in azimuth, $P_J(\phi)=(2\pi)^{-1}$. The conditional probability of observing a particle at an angle $\varphi_i$ relative to the jet direction is taken to be a delta function.

$$P_{PJ}(\varphi_i | \phi) = \delta(\varphi_i - \phi) \tag{3.1}$$

We further assume there exist definite probabilities, $P_J(J)$, and $P_J(A_i)$ for respectively observing J jets in a given event, and $A_i$ fragments associated with a given jet. The joint probability, $P(\varphi_i, \phi, J, A_i)$, of finding one of $A_i$ jet fragments at angle $\varphi_i$, from a jet emitted in direction $\phi$, while there are J jets in an event is given by:

$$P_J(\varphi_i, \phi, J, A_i) = P_{PJ}(\varphi_i | \phi) P_J(\phi) P_J(A_i) P_J(J) \tag{3.2}$$

We obtain the single particle density by integrating (marginalization[a]) unmeasured observables (e.g. in this case $\phi$) and averaging:

$$\rho_1(\varphi_i) = (2\pi)^{-1} \langle J \rangle \langle A_i \rangle, \tag{3.3}$$

where $\langle J \rangle$ and $\langle A_i \rangle$ are, respectively, the average number of jets in an event and the average number of particles associated with a jet. The two-particle density is similarly obtained by integrating the joint probability of finding jet fragments at angles $\varphi_i$, and $\varphi_j$, from a jet emitted in direction $\phi$, while there are J jets in an event. Note the two measured particles may be either from the same or different jets. The two-particle density is thus a sum of two terms as follows:

$$\rho_{2,J}(\varphi_i, \varphi_j) = \int J A_i A_j P_{PJ}(\varphi_i | \phi) P_{PJ}(\varphi_j | \phi) P_J(\phi) P_J(A_i, A_j) P_J(J) dJ dA_i dA_j d\phi$$
$$+ \int J(J-1) A_i A_j P_{PJ}(\varphi_i | \phi_\alpha) P_{PJ}(\varphi_j | \phi_\beta) P_J(\phi_\alpha) P_J(\phi_\beta) P_J(A_i, A_j) P_J(J) dJ dA_i dA_j d\phi_\alpha d\phi_\beta \tag{3.4}$$

The first term corresponds to the two-particle being emitted from the same jet, while the second term is for particles from two different jets $\alpha$, and $\beta$. Integration yields:

$$\rho_{2,J}(\varphi_i, \varphi_j) = (2\pi)^{-1} \langle J \rangle \langle A_i A_j \rangle P_{2,PJ}(\varphi_i, \varphi_j) + (2\pi)^{-2} \langle J(J-1) \rangle \langle A_i \rangle \langle A_j \rangle \tag{3.5}$$

where the probability $P_{2,PJ}(\varphi_i, \varphi_j)$ is given by

$$P_{2,PJ}(\varphi_i, \varphi_j) = \int_0^{2\pi} \delta(\varphi_i - \phi) \delta(\varphi_j - \phi) d\phi = \delta(\varphi_i - \varphi_j) \tag{3.6}$$

The 2-cumulant is obtained by applying the definition (2.2).

$$\hat{\rho}_{2,PJ}(\varphi_i, \varphi_j) = (2\pi)^{-1} \langle J \rangle \langle A_i A_j \rangle P_{2,PJ}(\varphi_i, \varphi_j) + (2\pi)^{-2} \langle A_i \rangle \langle A_j \rangle \left( \langle J(J-1) \rangle - \langle J \rangle^2 \right) \tag{3.7}$$

Note the second, constant term in the above expression vanishes if a Poissonian process determines the number of jets for which $\langle J(J-1) \rangle = \langle J \rangle^2$.

The three-particle density and 3-cumulant are obtained in a similar manner as the two-particle cumulant. It is possible for particles to be from the same, two, or three distinct jets. Integration of the multiplicity weighted joint probability yields:

---

[a] See definitions of joint probability and marginalization in Review of Particle Physics by the Particle Data Group.

$$\rho_{3,J}(\varphi_i,\varphi_j,\varphi_k) = (2\pi)^{-1}\langle J\rangle\langle A_iA_jA_k\rangle P_{3,PJ}(\varphi_i,\varphi_j,\varphi_k)$$
$$+(2\pi)^{-2}\langle J(J-1)\rangle\langle A_iA_j\rangle\langle A_k\rangle P_{2,PJ}(\varphi_i,\varphi_j)$$
$$+(2\pi)^{-2}\langle J(J-1)\rangle\langle A_iA_k\rangle\langle A_j\rangle P_{2,PJ}(\varphi_i,\varphi_k) \quad (3.8)$$
$$+(2\pi)^{-2}\langle J(J-1)\rangle\langle A_jA_k\rangle\langle A_i\rangle P_{2,PJ}(\varphi_j,\varphi_k)$$
$$+(2\pi)^{-3}\langle J(J-1)(J-2)\rangle\langle A_i\rangle\langle A_j\rangle\langle A_k\rangle$$

where the probability $P_{2,PJ}(\varphi_i,\varphi_j)$ is given by Eq. (3.6), and $P_{3,PJ}(\varphi_i,\varphi_j,\varphi_k)$ by

$$P_{3,PJ}(\varphi_i,\varphi_j,\varphi_k) = \int_0^{2\pi}\delta(\varphi_i-\phi)\delta(\varphi_j-\phi)\delta(\varphi_k-\phi)d\phi = \delta(\varphi_j-\varphi_i)\delta(\varphi_k-\varphi_i) \quad (3.9)$$

The 3-cumulant is.

$$\hat\rho_{3,PJ}(\varphi_i,\varphi_j,\varphi_k) = (2\pi)^{-1}\langle J\rangle\langle A_iA_jA_k\rangle P_{3,PJ}(\varphi_i-\varphi_j,\varphi_i-\varphi_k)$$
$$+(2\pi)^{-2}\left(\langle J(J-1)\rangle-\langle J\rangle^2\right)\begin{bmatrix}\langle A_iA_j\rangle\langle A_k\rangle P_{2,PJ}(\varphi_i-\varphi_j)\\+\langle A_iA_k\rangle\langle A_j\rangle P_{2,PJ}(\varphi_i-\varphi_j)\\+\langle A_jA_k\rangle\langle A_i\rangle P_{2,PJ}(\varphi_i-\varphi_j)\end{bmatrix} \quad (3.10)$$
$$+(2\pi)^{-3}\left(\langle J(J-1)(J-2)\rangle-3\langle J(J-1)\rangle\langle J\rangle+2\langle J\rangle^3\right)\langle A_i\rangle\langle A_j\rangle\langle A_k\rangle$$

We remark that if the number of jets, J, is determined by a Poissonian process, then one has $\langle J(J-1)\rangle=\langle J\rangle^2$, $\langle J(J-1)(J-2)\rangle-3\langle J(J-1)\rangle-2\langle J\rangle^3=0$ and the above expression therefore reduces to the following:

$$\hat\rho_{3,PJ}^{Poisson}(\varphi_i,\varphi_j,\varphi_k) = (2\pi)^{-1}\langle J\rangle\langle A_iA_jA_k\rangle\delta(\varphi_i-\varphi_j)\delta(\varphi_i-\varphi_k) \quad \mathbf{(3.11)}$$

We consider more realistic cases of jets and particle production, in the following sections. It is worth noting that while the specifics of the angular dependencies varies from case to case, cumulants will have similar expressions as in Eqs. 3.7 and 3.10 where appropriate angular distributions must be substituted to $P_{2,PJ}$ and $P_{3,PJ}$.

### 3.2. *Gaussian Jets (Model GJ)*

We consider the production of jets with a finite size (opening angle) and formulate the hypothesis that produced jets are not coupled to the bulk of produced particles. We can then separate the calculation of the jet and background correlations. We assume a given event consists of background particles, and jets. As in the previous section, we denote the probability distribution for finding "J" jets in a given event and $A_i$ associated particles by $P_J(J)$, and $P_J(A_i)$ respectively. We use a Gaussian profile to describe the fragment azimuthal distributions relative to the jet direction $\phi$. The conditional probability of observing a particle at angle $\varphi_i$ given the initial parton direction $\phi_\alpha$ is written:

$$P_J(\varphi_i|\phi_\alpha) = G(\varphi_i;\phi_\alpha,\sigma_i) \equiv \frac{1}{\sqrt{2\pi}\sigma_{i,\alpha}}\exp\left(-\frac{(\varphi_i-\phi_\alpha)^2}{2\sigma_i^2}\right) \quad (3.12)$$

For simplicity, we assume in this section the Gaussian widths are independent of the jet energy and other event attributes. We further assume the jets (partons) are emitted uniformly in azimuth, i.e., the probability of observing a jet at angle $\phi_\alpha$ is noted $P_J(\phi_\alpha)=(2\pi)^{-1}$. One finds the single particle density,

2- and 3- particle cumulants are given by expressions, (3.7), and (3.10) where the 2- and 3- particle probabilities are replaced by the following Gaussian profiles:

$$P_{2,GJ}(\varphi_1,\varphi_2) = \frac{1}{\sqrt{2\pi}\sigma_{1,2}}\exp\left(-\frac{(\varphi_1-\varphi_2)^2}{2\sigma_{1,2}^2}\right)$$

$$P_{3,GJ}(\varphi_1,\varphi_2,\varphi_3) = \frac{1}{2\pi\sigma_{1,2,3}^2}\exp\left(-\frac{\sigma_3(\varphi_1-\varphi_2)^2+\sigma_2(\varphi_1-\varphi_3)^2+\sigma_1(\varphi_2-\varphi_3)^2}{2\sigma_{1,2,3}^2}\right)$$

(3.13)

where $\sigma_{ij}^2 = \sigma_i^2 + \sigma_j^2$ and $\sigma_{i,j,k}^4 = \sigma_i^2\sigma_j^2 + \sigma_i^2\sigma_k^2 + \sigma_j^2\sigma_k^2$.

### 3.3. Anisotropic Flow (F)

Flow, or collective motion, is an important feature of heavy ion collisions at relativistic energies. It manifests itself by radial acceleration and modification of transverse momentum ($p_t$) spectra and by azimuthal anisotropy of produced particles. In this section, we focus our attention on azimuthal anisotropy arising in non-central heavy ion collisions relative to the reaction plane. It is convenient and customary to decompose the azimuthal anisotropy in terms of harmonics relative to an assumed reaction plane. The probability to observe a particle at a given azimuthal angle $\varphi_i$ relative to the reaction plane angle, $\psi$, is written as a Fourier series.

$$P_F(\varphi_i|\psi) = 1 + 2\sum_m v_m(i)\cos(m(\varphi_i-\psi))$$

(3.14)

The Fourier coefficients $v_m(i)$ measure the m-th order anisotropy for particles emitted in a selected kinematic range "i". Measurements have shown the second order (elliptical) anisotropy can be rather large in Au + Au collisions at RHIC while first and fourth order harmonics are typically much smaller. Little is know about third and fifth harmonics. Sixth harmonics have been estimated to be rather small at RHIC. STAR measurement shows the fourth harmonic scales roughly as the square of the 2dn harmonic ($v_4 \approx 1.1v_2^2$) [21,22,23]. Experimental techniques for measurements of flow harmonics are described at length in the literature. Our discussion here focuses on the impact of flow on two- and three-particle azimuthal correlations. We assume the harmonic coefficients are known. Measurements at a given collision centrality lead to average values for both the magnitude of the harmonic coefficients and the number of produced particles. For the purpose of this model, we describe the probability of finding $F_i$ particles in the kinematical range "i" according to probability $P(F_i)$. The exact form of this probability is not required. Only the first, second, and third moments are needed. The joint probability of measuring $F_i$, at an angle $\varphi_i$ while the reaction plane angle is at $\psi$ is given by:

$$P_F(\varphi_i,F_i,\psi) = P_F(\varphi_i|\psi)P_F(F_i)P_F(\psi)$$

(3.15)

where $P(\psi)=(2\pi)^{-1}$ is the probability of finding the reaction plane at a given angle $\psi$. Integration yields the single particle density $\rho_1(\varphi_i) = (2\pi)^{-1}\langle F_i\rangle$. The probability to observe two particles at angles $\varphi_i$, and $\varphi_j$ may be written:

$$P_F(\varphi_i,F_i,\varphi_j,F_j,\psi) = P_F(\varphi_i|\psi)P_F(\varphi_j|\psi)P(F_i,F_j)P_F(\psi)$$

(3.16)

Integration of this probability yields the 2-particle density:

$$\rho_2(\varphi_i,\varphi_j) = (2\pi)^{-1}\langle F_iF_j\rangle P_{2,F}(\varphi_i,\varphi_j)$$

(3.17)

where $P_{2,F}(\varphi_i, \varphi_j)$ is given by:

$$P_{2,F}(\varphi_i,\varphi_j) = 1 + 2\sum_m v_m(i)v_m(j)\cos(m(\varphi_i-\varphi_j))$$

(3.18)

Similarly, one finds the three-particle density is:

$$\rho_3(\varphi_i,\varphi_j,\varphi_k) = (2\pi)^{-3} \langle F_i F_j F_k \rangle P_{3,F}(\varphi_i,\varphi_j,\varphi_k) \tag{3.19}$$

where $P_F(\varphi_i, \varphi_j, \varphi_k)$ is given by the following expression:

$$P_{3,F}(\varphi_i,\varphi_j,\varphi_k) = 1 + 2\sum_m v_m(i)v_m(j)\cos(m(\varphi_i-\varphi_j)) + 2\sum_m v_m(i)v_m(k)\cos(m(\varphi_i-\varphi_k))$$

$$+ 2\sum_m v_m(j)v_m(k)\cos(m(\varphi_j-\varphi_k)) + 2\sum_{p,m,n} v_p(i)v_m(j)v_n(k)\begin{bmatrix} \delta_{p,m+n}\cos(p\varphi_i - m\varphi_j - n\varphi_k) \\ +\delta_{m,p+n}\cos(-p\varphi_i + m\varphi_j - n\varphi_k) \\ +\delta_{n,m+k}\cos(-p\varphi_i - m\varphi_j + n\varphi_k) \end{bmatrix} \tag{3.20}$$

The 2- and 3-cumulants are as follows:

$$\hat{\rho}_{2,F}(\varphi_i,\varphi_j) = (2\pi)^{-2}\left( \langle F_i F_j \rangle - \langle F_i \rangle \langle F_j \rangle + 2\langle F_i F_j \rangle \sum_m v_m(i)v_m(j)\cos(m(\varphi_i-\varphi_j)) \right) \tag{3.21}$$

$$\hat{\rho}_{3,F}(\varphi_i,\varphi_j,\varphi_k) = (2\pi)^{-3}\left\{ \begin{array}{l} 2\langle F_i F_j F_k \rangle \sum_{p,m,n} v_p(i)v_m(j)v_n(k)\begin{bmatrix} \delta_{p,m+n}\cos(p\varphi_i - m\varphi_j - n\varphi_k) \\ +\delta_{m,p+n}\cos(-p\varphi_i + m\varphi_j - n\varphi_k) \\ +\delta_{n,m+k}\cos(-p\varphi_i - m\varphi_j + n\varphi_k) \end{bmatrix} \\ +2(\langle F_i F_j F_k \rangle - \langle F_i F_j \rangle \langle F_k \rangle)\sum_m v_m(i)v_m(j)\cos(m(\varphi_i-\varphi_j)) \\ +2(\langle F_i F_j F_k \rangle - \langle F_i F_k \rangle \langle F_j \rangle)\sum_m v_m(i)v_m(k)\cos(m(\varphi_i-\varphi_k)) \\ +2(\langle F_i F_j F_k \rangle - \langle F_j F_k \rangle \langle F_i \rangle)\sum_m v_m(j)v_m(k)\cos(m(\varphi_j-\varphi_k)) \\ -\langle F_i F_j \rangle \langle F_k \rangle - \langle F_i F_k \rangle \langle F_j \rangle - \langle F_j F_k \rangle \langle F_i \rangle + 2\langle F_i \rangle \langle F_j \rangle \langle F_k \rangle \end{array} \right\} \tag{3.22}$$

Note that the above expression contains both second and third order terms in $v_m$. However, if the particle production process is Poissonian, one has $\langle F_i F_j F_k \rangle = \langle F_i F_j \rangle \langle F_k \rangle$, and $\langle F_i F_j \rangle = \langle F_i \rangle \langle F_j \rangle$ so the above expression reduces to:

$$\hat{\rho}_{3,F}^{Poissonian}(\varphi_i,\varphi_j,\varphi_k) = 2(2\pi)^{-3}\langle F_i \rangle \langle F_j \rangle \langle F_k \rangle \sum_{p,m,n} v_p(i)v_m(j)v_n(k)\begin{bmatrix} \delta_{p,m+n}\cos(p\varphi_i - m\varphi_j - n\varphi_k) \\ +\delta_{m,p+n}\cos(-p\varphi_i + m\varphi_j - n\varphi_k) \\ +\delta_{n,m+k}\cos(-p\varphi_i - m\varphi_j + n\varphi_k) \end{bmatrix} \tag{3.23}$$

which has no second order term in harmonic $v_m$, and only non-diagonal terms in $v_m(i)v_n(j)v_p(k)$. While the above cumulant is reduced to a rather simple form for Poissonian processes, it is generally not warranted to assume the production processes are Poissonian. Indeed, measurements and comparisons of $\langle N_i N_j N_k \rangle$, $\langle N_i N_j \rangle \langle N_k \rangle$, and $\langle N_i \rangle \langle N_j \rangle \langle N_k \rangle$ reveal these quantities are generally different. That implies the above 3-cumulant shall, in fact, have finite flow harmonics of all orders. Note that one could, in principle, "define" flow as being a Poissonian process thereby reducing (3.22) to the above expression by definition. Consider, however, that realistic modeling of collisions for the purpose of extracting jet and Mach cone signals would then require one also include additional terms in the model to account for non-flow contributions arising from resonance decay, electric (strangeness, baryon number) charge conservation, etc. It is thus convenient to assume the harmonic anisotropies to be non-Poissonian and subsume all non-flow effects (other than jet + Mach cone) into harmonic anisotropies as discussed here. This approach, however, implies one is seemingly stuck with $v_m^2$ contributions. This is particularly disappointing because the use of 3-cumulants for jet measurements is, in part, predicated by the notion that these second orders can be eliminated in a straightforward manner. There is, however, an alternative solution to this problem. The solution resides in a modest modification of the cumulants' definition. Indeed, since the usual cumulants lead to the presence of irreducible terms in $\langle F_i F_j F_k \rangle - \langle F_i F_j \rangle \langle F_k \rangle$, it is natural to define modified 3-cumulants as follows:

$$\hat{\rho}_3(\varphi_i,\varphi_j,\varphi_k) = \frac{\langle F_i \rangle \langle F_j \rangle \langle F_k \rangle}{\langle F_i F_j F_k \rangle} \rho_3(\varphi_i,\varphi_j,\varphi_k)$$

$$- \frac{\langle F_i \rangle \langle F_j \rangle}{\langle F_i F_j \rangle} \rho_2(\varphi_i,\varphi_j)\rho_1(\varphi_k) - \frac{\langle F_i \rangle \langle F_k \rangle}{\langle F_i F_k \rangle} \rho_2(\varphi_i,\varphi_k)\rho_1(\varphi_j) - \frac{\langle F_j \rangle \langle F_k \rangle}{\langle F_j F_k \rangle} \rho_2(\varphi_j,\varphi_k)\rho_1(\varphi_i) \quad (3.24)$$

$$+ 2\rho_1(\varphi_i)\rho_1(\varphi_j)\rho_1(\varphi_k)$$

It is straightforward to show this modified cumulant leads to the same result as the Poissonian hypothesis.

$$\hat{\rho}_{3,F}(\varphi_i,\varphi_j,\varphi_k) = 2(2\pi)^{-3} \langle F_i \rangle \langle F_j \rangle \langle F_k \rangle \sum_{p,m,n} v_p(i) v_m(j) v_n(k) \begin{bmatrix} \delta_{p,m+n} \cos(p\varphi_i - m\varphi_j - n\varphi_k) \\ +\delta_{m,p+n} \cos(-p\varphi_i + m\varphi_j - n\varphi_k) \\ +\delta_{n,m+k} \cos(-p\varphi_i - m\varphi_j + n\varphi_k) \end{bmatrix} \quad (3.25)$$

In the above expression, the Kronecker deltas imply that only non-diagonal terms contribute to the cumulant. Examples of such terms are $v_1(i)v_1(j)v_2(k)$, $v_1(i)v_2(j)v_3(k)$, $v_2(i)v_2(j)v_4(k)$, etc. Coefficients $v_1$ measured at RHIC are rather small[21,22,23], on the order of 0.01, while $v_2$ coefficients can be as large as 0.2. $v_4$ coefficients have been measured[22] to be on the order of $v_2^2$. Little is known about $v_3$ coefficients, but on general grounds, one can expect them to be of the same order or smaller than $v_1$ coefficients. One therefore expects the leading terms in (3.22) should be the $v_2(i)v_2(j)v_4(k)$ terms. These are shown in

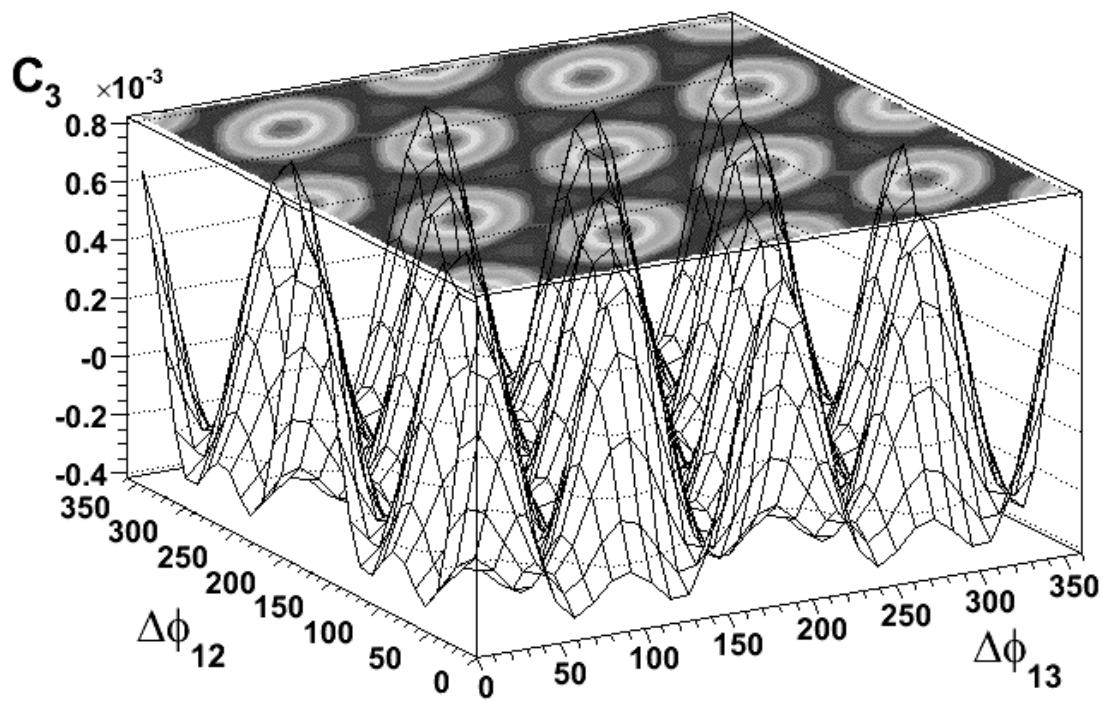

Figure 2 with equal coefficients for illustrative purposes. In general, the values of the $v_2$ and $v_4$ coefficients depend on the kinematic ranges i, j, and k, used to calculated them. There is, therefore, no reason to expect that the three terms inside the sum should contribute equally.

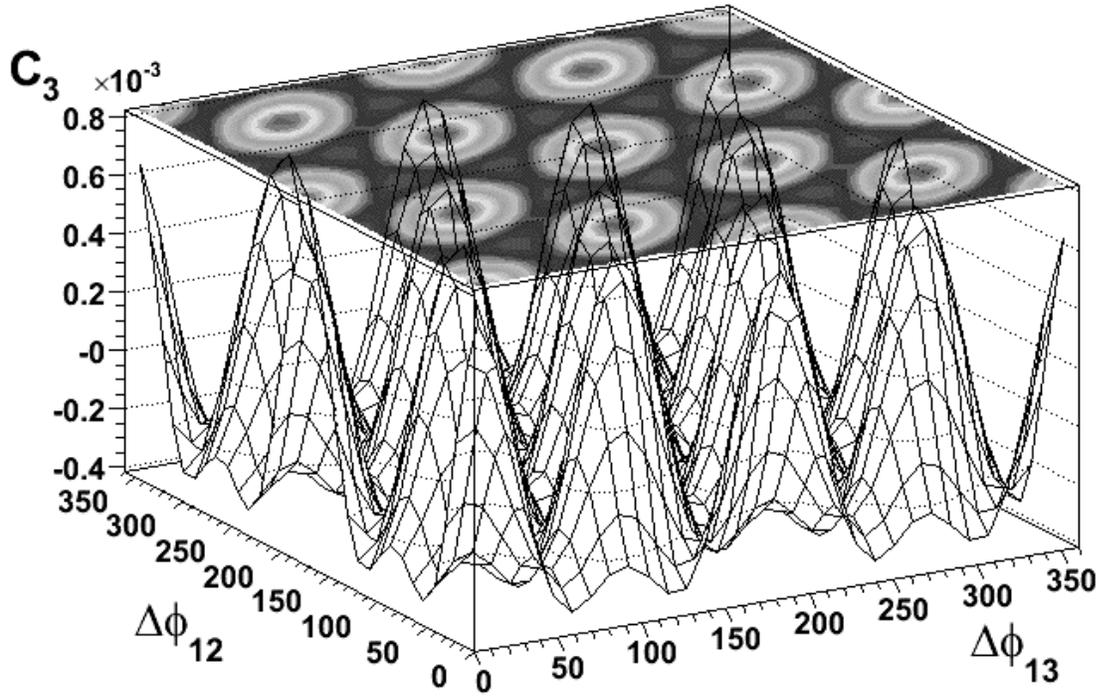

Figure 2 Contour plot of the 3-cumulant (3.24) for finite $v_2v_2v_4$ harmonic flow calculated with equal arbitrary coefficients.

Given the expression (3.24), one expects that the irreducible flow harmonics can produce a rather intricate and non-trivial shape that may partly mock-up or mask the signal expected from proposed signals for Mach cone, and Cerenkov gluon radiation. It is thus essential to understand and control the magnitude of such terms before attempting to identify these proposed exotic new phenomena.

Whether the intricate non-diagonal anisotropies must be explicitly accounted for depends on their strength relative to the jet 3-cumulant signals. We evaluated the magnitude of the flow harmonic coefficients on the basis of a parameterization of data published by STAR[24]. As a practical example, we present an estimate of the $v_2v_2v_4$ contributions for trigger particle (1) with $4 < p_t < 10$ GeV/c, $|\eta|<1$, and associated particles (2,3) with $0.15 < p_t < 4$ GeV/c, $|\eta|<1$ in Table 1. The amplitude of the $v_2v_2v_4$ term is largest in peripheral collisions and becomes progressively smaller for central collisions. While amplitudes shown in Table 1 are indeed rather small, one must consider that the strength of correlations seen in three-particle are also modest, i.e. on the order $10^{-3}$ to $10^{-4}$. The $v_2v_2v_4$ terms are thus expected to form a sizeable background in three-particle cumulant analyses.

Table 1. Estimate of the amplitude of v2v2v4 non-diagonal terms based on published $v_2$ data.

| Centrality | $v_2(1)$ | $v_2(2), v_2(3)$ | $v_2(1)v_2(2)v_2(3)^2$ | $v_2(1)^2v_2(2)v_2(3)$ |
|---|---|---|---|---|
| 1 | 0.193 | 0.074 | 7.7 x$10^{-5}$ | 2.0x$10^{-4}$ |
| 2 | 0.237 | 0.092 | 1.8 x$10^{-4}$ | 4.7 x$10^{-4}$ |
| 3 | 0.197 | 0.095 | 1.7x$10^{-4}$ | 3.5x$10^{-4}$ |
| 4 | 0.193 | 0.097 | 1.8x$10^{-4}$ | 3.5x$10^{-4}$ |
| 5 | 0.180 | 0.094 | 1.5x$10^{-4}$ | 2.9x$10^{-4}$ |
| 6 | 0.175 | 0.083 | 9.9x$10^{-5}$ | 2.1x$10^{-4}$ |
| 10-20% | 0.133 | 0.064 | 3.5x$10^{-5}$ | 7.3x$10^{-5}$ |
| 5-10% | 0.094 | 0.044 | 7.8x$10^{-6}$ | 1.7x$10^{-5}$ |
| 0-5% | 0.048 | 0.025 | 7.7x$10^{-7}$ | 1.5x$10^{-6}$ |

### 3.4. *Gaussian Jets + Harmonic Flow (Model GJ+F)*

We explore the case of a superposition of jets and harmonic flow treated as statistically independent processes. Many recent analyses of RHIC data essentially reduce to such a case. As stated in Section 2, the cumulant of a sum of independent processes is equal to the sum of the cumulants of each of the processes. It is thus trivial to write the single particle density, 2- and 3-cumulants for this particular model, based on results obtained in previous sections.

The single particle density is simply:

(3.26) $\rho(\varphi_i) = (2\pi)^{-1} \left( \langle J \rangle \langle A_i \rangle + \langle F_i \rangle \right)$

where $\langle J \rangle$, $\langle A_i \rangle$, and $\langle F_i \rangle$ respectively denote the average number of jets, the average number of associated jet fragments per jet, and the average number of particles in the "flow background" for the given kinematical range "i". The 2-cumulant is readily found to be:

$$\hat{\rho}_{2,J+F}(\varphi_i, \varphi_j) = (2\pi)^{-1} \langle J \rangle \langle A_i \rangle \langle A_j \rangle P_{2,G}(\varphi_i, \varphi_j; \sigma_i, \sigma_j)$$

$$+ (2\pi)^{-2} \langle A_i \rangle \langle A_j \rangle \left( \langle J(J-1) \rangle - \langle J \rangle^2 \right) \qquad (3.27)$$

$$+ (2\pi)^{-2} \left[ \langle F_i F_j \rangle - \langle F_i \rangle \langle F_j \rangle + 2 \langle F_i F_j \rangle \sum_m v_m(i) v_m(j) \cos\left( m(\varphi_i - \varphi_j) \right) \right]$$

where $P_{2,G}$ given by (3.12) corresponds to the assumed jet Gaussian profile with width $\sqrt{\sigma_i^2 + \sigma_j^2}$. The last term contains flow harmonics dominated by the second harmonic, although it is also recognized that fourth harmonics may also play a finite role. The 3-cumulant is given by:

$$\hat{\rho}_{3,J+F}(\varphi_i, \varphi_j, \varphi_k) = (2\pi)^{-1} \langle J \rangle \langle A_i A_j A_k \rangle P_{3,G}(\varphi_i, \varphi_j, \varphi_k; \sigma_i, \sigma_j, \sigma_k)$$

$$+ (2\pi)^{-2} \left( \langle J(J-1) \rangle - \langle J \rangle^2 \right) \langle A_i A_j \rangle \langle A_k \rangle P_{2,G}(\varphi_i, \varphi_j; \sigma_i, \sigma_j)$$

$$+ (2\pi)^{-2} \left( \langle J(J-1) \rangle - \langle J \rangle^2 \right) \langle A_i A_k \rangle \langle A_j \rangle P_{2,G}(\varphi_i, \varphi_j; \sigma_i, \sigma_k)$$

$$+ (2\pi)^{-2} \left( \langle J(J-1) \rangle - \langle J \rangle^2 \right) \langle A_j A_k \rangle \langle A_i \rangle P_{2,G}(\varphi_j, \varphi_k; \sigma_j, \sigma_k)$$

$$+ (2\pi)^{-3} \left( \langle J(J-1)(J-2) \rangle - 3\langle J \rangle \langle J(J-1) \rangle + 2\langle J \rangle^2 \right) \langle A_k \rangle \langle A_j \rangle \langle A_i \rangle$$

$$(2\pi)^{-3} \left\{ \begin{array}{l} 2\langle F_i F_j F_k \rangle \sum_{p,m,n} v_p(i) v_m(j) v_n(k) \left[ \begin{array}{l} \delta_{p,m+n} \cos(p\varphi_i - m\varphi_j - n\varphi_k) \\ + \delta_{m,p+n} \cos(-p\varphi_i + m\varphi_j - n\varphi_k) \\ + \delta_{n,m+k} \cos(-p\varphi_i - m\varphi_j + n\varphi_k) \end{array} \right] \\ + 2\left( \langle F_i F_j F_k \rangle - \langle F_i F_j \rangle \langle F_k \rangle \right) \sum_m v_m(i) v_m(j) \cos\left( m(\varphi_i - \varphi_j) \right) \\ + 2\left( \langle F_i F_j F_k \rangle - \langle F_i F_k \rangle \langle F_j \rangle \right) \sum_m v_m(i) v_m(k) \cos\left( m(\varphi_i - \varphi_k) \right) \\ + 2\left( \langle F_i F_j F_k \rangle - \langle F_j F_k \rangle \langle F_i \rangle \right) \sum_m v_m(j) v_m(k) \cos\left( m(\varphi_j - \varphi_k) \right) \\ - \langle F_i F_j \rangle \langle F_k \rangle - \langle F_i F_k \rangle \langle F_i \rangle - \langle F_j F_k \rangle \langle F_i \rangle + 2\langle F_i \rangle \langle F_j \rangle \langle F_k \rangle \end{array} \right\} \qquad (3.28)$$

While the above cumulant reduces to a rather simple form for Poissonian processes, it is generally not warranted to assume the production processes are Poissonian. Indeed, measurements and comparisons of $\langle N_i N_j N_k \rangle$, $\langle N_i N_j \rangle \langle N_k \rangle$, and $\langle N_i \rangle \langle N_j \rangle \langle N_k \rangle$ reveal these quantities are generally different. That implies the above 3-cumulant shall, in fact, have finite flow harmonics of all orders. It is, however, possible in principle, to partly mitigate this problem if one can use $\langle N_i N_j N_k \rangle$ and $\langle N_i N_j \rangle$ as estimators of $\langle F_i F_j F_k \rangle$ and $\langle F_i F_j \rangle$, respectively, and use the modified 3-cumulant definition (3.24). One then gets an expression where only non-diagonal, high order, harmonic components contribute. Defining normalization coefficients $\alpha$ and $\beta_{ij}$ as follows:

$$\alpha = \frac{\langle N_i \rangle \langle N_j \rangle \langle N_k \rangle}{\langle N_i N_j N_k \rangle}$$

$$\beta_{ij} = \frac{\langle N_i \rangle \langle N_j \rangle}{\langle N_i N_j \rangle}$$

(3.29)

One obtains the following expression for the 3-cumulant:

$$\hat{\rho}_{3,J+F}(\varphi_i,\varphi_j,\varphi_k) = (2\pi)^{-1} \alpha \langle J \rangle \langle A_i A_j A_k \rangle P_{3,G}(\varphi_i,\varphi_j,\varphi_k;\sigma_i,\sigma_j,\sigma_k)$$
$$+ (2\pi)^{-2} \left( \alpha \langle J(J-1) \rangle - \beta_{ij} \langle J \rangle^2 \right) \langle A_i A_j \rangle \langle A_k \rangle P_{2,G}(\varphi_i,\varphi_j;\sigma_i,\sigma_j)$$
$$+ (2\pi)^{-2} \left( \alpha \langle J(J-1) \rangle - \beta_{ij} \langle J \rangle^2 \right) \langle A_i A_k \rangle \langle A_j \rangle P_{2,G}(\varphi_i,\varphi_k;\sigma_i,\sigma_k)$$
$$+ (2\pi)^{-2} \left( \alpha \langle J(J-1) \rangle - \beta_{ij} \langle J \rangle^2 \right) \langle A_j A_k \rangle \langle A_i \rangle P_{2,G}(\varphi_j,\varphi_k;\sigma_j,\sigma_k)$$
$$+ (2\pi)^{-3} \begin{pmatrix} \alpha \langle J(J-1)(J-2) \rangle \\ -(\beta_{12} + \beta_{13} + \beta_{23}) \langle J \rangle \langle J(J-1) \rangle \\ +(\beta_{12} + \beta_{13} + \beta_{23} - 1) \langle J \rangle^2 \end{pmatrix} \langle A_k \rangle \langle A_j \rangle \langle A_i \rangle$$
$$(2\pi)^{-3} 2 \langle F_i \rangle \langle F_j \rangle \langle F_k \rangle \sum_{p,m,n} v_p(i) v_m(j) v_n(k) \begin{bmatrix} \delta_{p,m+n} \cos(p\varphi_i - m\varphi_j - n\varphi_k) \\ +\delta_{m,p+n} \cos(-p\varphi_i + m\varphi_j - n\varphi_k) \\ +\delta_{n,m+k} \cos(-p\varphi_i - m\varphi_j + n\varphi_k) \end{bmatrix}$$

(3.30)

We note this simpler expression is only valid if the estimator $\langle N_i N_j N_k \rangle$ and $\langle N_i N_j \rangle$ exactly equal $\langle F_i F_j F_k \rangle$ and $\langle F_i F_j \rangle$, respectively. If a perfect match cannot be accomplished, there shall be some finite residual $v_2(i)v_2(j)$ harmonic components.

### 3.5. *Gaussian Di-Jets, and Scattered Jets (Model SJ)*

We now seek a simple representation of di-jets to investigate whether jet scattering effects can be properly disentangled in a measurement based on two- and three-particle azimuthal correlations. Given that cumulants corresponding to a sum of processes can be written as the sum of the cumulant of each processes, we will restrict the discussion in this section to scattering effects. Obviously, addition of flow terms or other types of uncorrelated processes, can be added as the model presented in the previous section. This assumption, however, becomes invalid if the jet emission is modulated in azimuth, relative to the reaction plane, as discussed in Section 3.7.

We model the two jets of a di-jet with Gaussian distributions centered at angles ϕ and ϕ+Δϕ respectively. Effectively, one integrates jets scattered at forward/backward rapidities and neglects corrections of order $(\cos\Delta\eta)^{-1}$ in the expression of the cross-section. The joint probability distribution is taken as:

$$P_{DJ}(\varphi_i, A_i, \phi, \Delta\phi, J,) = \sum_{r=0}^{1} P_G(\varphi_i; \sigma_{i,r}, \phi + r\Delta\phi) P_J(A_i) P_J(J) P_J(\phi) \left( \delta_{0,r} + \delta_{1,r} P_S(\Delta\phi; \Delta\phi_o, \sigma_{\Delta\phi}) \right) \quad (3.31)$$

The label r=0 is used to denote the trigger or leading jet while r=1 is used for the away-side jet. $P_G$ is a Gaussian probability distribution expressing the probability of finding a fragment at angle $\varphi_i$ relative to the jet direction ϕ. $P_J(A_{i,r})$ expresses the probability of finding $A_{i,r}$ fragments associated with the lead (r=0) and away-side (r=1) jets, which are assigned widths $\sigma_{i,r}$. $P_J(J)$ is the probability of the number of di-jet J. $P_J(\phi)) = (2\pi)^{-1}$ is the probability of finding the lead jet axis at an angle ϕ. The direction of the away-side jet is determined by the angle Δϕ. Very high $p_t$ jets should be produced essentially back-to-back in azimuth, i.e. with Δϕ ~ 180°. At lower $p_t$, the relative angle may deviate significantly from 180° on an event-by-event basis while the jet widths themselves do not actually change. This can be modeled by assigning a scattering probability $P_S(\Delta\phi)$ to each value Δϕ. For simplicity, we use a Gaussian distribution with mean, $\langle \Delta\phi \rangle = \Delta\phi_o = 180^\circ$, and width, $\sigma_{\Delta\phi}$, to describe the scattering in the above expression. The same function can also be used to represent a Mach cone type effect if one neglects forward/backward emission of gluons by setting the scattering angle to ~120°. Indeed, according to Shuryak *et al.*, the emission angle of the Mach cone is determined by the speed of the away-side parton relative to the speed of sound, which they estimate is on the order of 0.33 *c*.

The single particle density corresponding to this model is obtained by integrating the joint probability multiplied by J, and $A_{i,r}$ over all non-observed variables. One gets:

$$\rho_1(\varphi_i) = \langle J \rangle \sum_{r=0}^{1} \langle A_{i,r} \rangle \quad (3.32)$$

Derivation of the two- and three-particle densities and cumulants proceeds similarly to that of model GJ discussed in sections 3.2 and 3.4. One must account, however, for the fact we have included a Gaussian dispersion with width $\sigma_\phi$ for the relative angle Δϕ. One gets the 2-cumulant:

$$\hat{\rho}_{2,DJ}(\varphi_i, \varphi_j; \sigma_{i,r}, \sigma_{j,s}, \Delta\phi_o, \sigma_{\Delta\phi}) = (2\pi)^{-1} \langle J \rangle \sum_{r,s=0}^{1} \langle A_{i,r} A_{j,s} \rangle P_{2GJ'}(\varphi_i, \varphi_j; \sigma_{i,r}, \sigma_{j,s}, (r-s)\Delta\phi_o, |r-s|\sigma_{\Delta\phi})$$
$$+ (2\pi)^{-2} \left( \langle J(J-1) \rangle - \langle J \rangle^2 \right) \sum_{r,s=0}^{1} \langle A_{i,r} \rangle \langle A_{j,s} \rangle \quad (3.33)$$

$P_{2,GJ'}$ is a generalization of (3.24) which includes $\sigma_\phi$ scattering effects. It is found to be:

$$P_{2,GJ'}(\varphi_i, \varphi_j; \sigma_i, \sigma_j, \Delta\phi, \sigma_\phi) = \frac{1}{\sqrt{2\pi}\sigma'} \exp\left( -\frac{(\varphi_i - \varphi_j - \Delta\phi)^2}{2\sigma'^2} \right) \quad (3.34)$$

where $\sigma'^2 = \sigma_i^2 + \sigma_j^2 + \sigma_\phi^2$. Note that for $r \neq s$, the associated multiplicities are likely to be uncorrelated and one should then have $\langle A_{i,r} A_{j,s} \rangle = \langle A_{i,r} \rangle \langle A_{j,s} \rangle$ for $r \neq s$.

The calculation of the 3-cumulant thus yields:

$$\hat{\rho}_{3,J}(\varphi_i,\varphi_j,\varphi_k;\sigma_{i,r},\sigma_{j,s},\sigma_{k,t},\Delta\phi_o,\sigma_\phi)$$

$$= (2\pi)^{-1}\langle J\rangle \sum_{r,s,t=0}^{1} \langle A_{i,r}A_{j,s}A_{k,t}\rangle P_{3,GJ'}(\varphi_i,\varphi_j,\varphi_k;r,s,t,\sigma_{i,r},\sigma_{j,s},\sigma_{k,t},\Delta\phi_o,\sigma_\phi)$$

$$+(2\pi)^{-2}\left(\langle J(J-1)\rangle - \langle J\rangle^2\right)\left\{\begin{array}{l}\sum_{r,s,t=0}^{1}\langle A_{i,r}A_{j,s}\rangle\langle A_{k,t}\rangle P_{2,GJ'}(\varphi_i,\varphi_j;\sigma_{i,r},\sigma_{j,s},(r-s)\Delta\phi_o,|r-s|\sigma_\phi) \\ +\sum_{r,s,t=0}^{1}\langle A_{i,r}A_{k,t}\rangle\langle A_{j,s}\rangle P_{2,GJ'}(\varphi_i,\varphi_k;\sigma_{i,r},\sigma_{k,t},(r-t)\Delta\phi_o,|r-t|\sigma_\phi) \\ +\sum_{r,s,t=0}^{1}\langle A_{j,s}A_{k,t}\rangle\langle A_{i,r}\rangle P_{2,GJ'}(\varphi_j,\varphi_k;\sigma_{j,s},\sigma_{k,t},(s-t)\Delta\phi_o,|s-t|\sigma_\phi)\end{array}\right\} \quad (3.35)$$

$$+(2\pi)^{-3}\left(\langle J(J-1)(J-2)\rangle - 3\langle J(J-1)\rangle\langle J\rangle + 2\langle J\rangle^3\right)\sum_{r,s,t=0}^{1}\langle A_{i,r}\rangle\langle A_{j,s}\rangle\langle A_{k,t}\rangle$$

The function $P_{3GJ'}$ is a generalized version of (3.13), which includes jet-scattering effects.

$$P_{3G'}(\varphi_i,\varphi_j,\varphi_k;r,s,t,\Delta\phi_0,\sigma_\phi) = \frac{1}{2\pi\sigma_{3G'}}\exp\left[-\frac{1}{2\sigma_{3G'}^4}\left(\begin{array}{l}\sigma_k^2\left((r-s)\Delta\phi_0 - \Delta\varphi_{ij}\right)^2 \\ +\sigma_j^2\left((r-t)\Delta\phi_0 - \Delta\varphi_{ik}\right)^2 \\ +\sigma_i^2\left((s-t)\Delta\phi_0 - \Delta\varphi_{jk}\right)^2 \\ +\sigma_\phi^2\left((t-s)\varphi_i + (r-t)\varphi_j(s-r)\varphi_k\right)^2\end{array}\right)\right] \quad (3.36)$$

Note that, as in Sect 3.2, the 3-cumulant contains 2-body terms (e.g. proportional to $P_{2,GJ'}$), which vanish for Poissonian jet production. The "intrinsic" width of the jets is effectively increased by the width of the scattering function. We illustrate this effect in Figure 3 where we compare calculations of $\Delta\varphi_{12}$ vs. $\Delta\varphi_{13}$ 3-particle correlations calculated with the above equation using equal distribution widths of 10° for the three particles. The correlation shown on the left includes no away-side parton deflection, whereas the correlation shown on the right was calculated with 30° for the width of the deflection function, and 180° for the mean deflection angle $\Delta\varphi_0$. The deflection broadens and reduces the relative amplitude of the away-side peaks. Note that parton energy loss should contribute an additional reduction of the away-side amplitude. The examples shown in Figure 3 were calculated with equal Gaussian widths for all three particles. In practice, one finds that high pt particles are characterized by smaller widths than low pt particles. Also note that formula (3.36) neglects parton scattering in the forward/backward direction. Including this scattering effectively results in a 1/cos(Δφ), which should have a modest impact on the shape of the distribution except for very broad away-side jet peaks. Finally, note that jet interaction with the medium shall contribute additional broadening of the away-side jet (and perhaps also the near side jet). Away-side jet broadening should be visible along both the main and secondary diagonal of the three-particle cumulant shown in Figure 3.

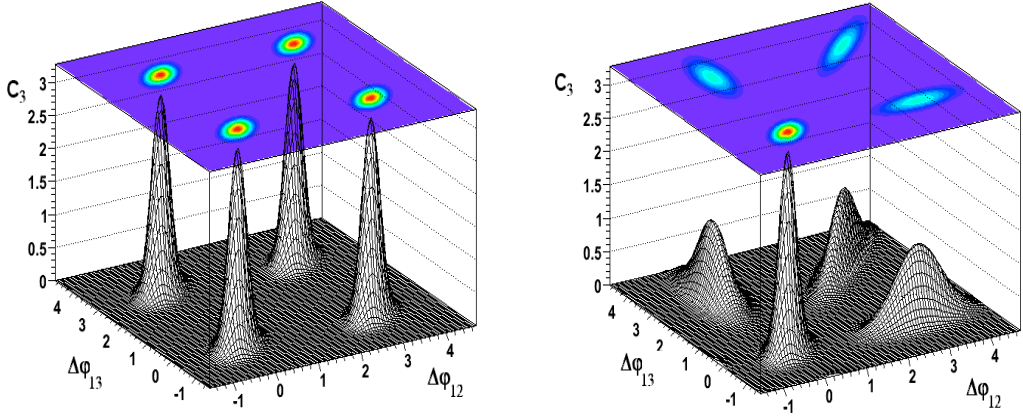

Figure 3. Three-particle cumulants expected for Gaussian jets modeled with Eq.(3.35). (Left) Particles produced with Gaussian distributions of 10° width but no scattering or deflection of the away-side parton (jet). (Right) Random azimuthal deflection of the away-side parton estimated with a Gaussian distribution of 30° width (in color online).

### 3.6. Two-body Decay (TBD)

We next consider the decay of resonances such as $\rho^0$ and $\Delta^0$ with a simple thermal model to illustrate that such decay may lead to non-trivial structures in two-particle correlations, while producing null signal in three-particle cumulants. We show that the structures produced in two-particle correlations are typically non-Gaussian and have shapes which depend on the specific transverse momentum and rapidity ranges considered, as well as complicated dependence on the parameters of the model, i.e., temperature and radial boost velocity. This example also provides a crude model of correlations induced by di-jet flow: production of di-jet fragments which experience large radial boost because of interaction with the medium.

Our calculation is based on a particle production model inspired from the blast-wave model. The resonances are produced to have a transverse momentum spectrum determined by:

$$E\frac{d^3N}{dp^3} = ke^{-\beta(E-\vec{v}_\perp \cdot \vec{p})} \tag{3.37}$$

where $\vec{v}_\perp$ is a transverse velocity boost. For simplicity, both $\vec{v}_\perp$ and the inverse temperature $\beta$ are a constant in the simulations shown.

Like the case of pencil-like jets (see section 3.1), one writes the two-particle cumulant associated with two-body decays (TBD) as:

$$\hat{\rho}_{2,TBD}(\varphi_i,\varphi_j) = (2\pi)^{-1}\langle N\rangle\langle A_iA_j\rangle P_{2,TBD}(\varphi_i,\varphi_j) + (2\pi)^{-2}\left(\langle N(N-1)\rangle - \langle N\rangle^2\right)\langle A_i\rangle\langle A_j\rangle \tag{3.38}$$

In this expression, the function $P_{2,TBD}(\varphi_i, \varphi_j)$ represents the probability of measuring the decays' products at the given angles. $\langle N\rangle$ represents the average number of decaying resonances while $\langle A_i\rangle$ and $\langle A_j\rangle$ are the average single particle yields resulting from the decays in the kinematic ranges "i" and "j". Because the emission of decay products is correlated and constrained by momentum and energy conservation, the product $\langle A_iA_j\rangle$ (which represents the average number of decays pairs detected simultaneously) shall, in general, be much smaller than the product $\langle A_i\rangle \langle A_j\rangle$. Although $\langle A_iA_j\rangle$, and $P_{2,TBD}(\varphi_i, \varphi_j)$ can be evaluated analytically for some kinematic ranges, it is more convenient, in this work, to perform a computation using simple Monte Carlo generators. We simulated the production of $\pi^+$, $\pi^-$, and $\rho^0$ at fixed temperature and radial velocity using Eq. 3.37. The relative abundance of the three species was set event-by-event using a multinomial generator with an average number of $\pi^+$, $\pi^-$ $p_1$ and $p_2$ and an average number of $\rho^0$ equal to 1-$p_1$-$p_2$. The number of particles ($\pi^+$, $\pi^-$, or $\rho^0$) was randomly generated with a flat distribution between 40 and 50 particles.

We begin with a discussion of rho-decays in the context of two-particle correlations and show how the kinematic ranges selected for analysis may influence the shape of the correlation function. We conclude this section with a discussion of three-particle correlation and show that, while the three-particle density exhibits finite structures, the three-particle cumulant is featureless.

Figure 4 displays two-particle correlation functions obtained with the $\rho^0$ decay toy model described above. The temperature of the pions is set to 0.4 GeV. No radial flow is used. For illustrative purposes, the $\rho^0$ are produced with selected rho-meson transverse momentum ranges and analysis cuts as follows: (a) $0.01 < p_t(\rho^o) < 0.1$ GeV/c, $p_t(1,2) < 0.2$ GeV/c; (b) $0.1 < p_t(\rho^o) < 0.5$ GeV/c, $p_t(1) > 0.3$ GeV/c, $p_t(2) < 0.2$ GeV/c (c) $0.1 < p_t(\rho^o) < 0.5$ GeV/c, $p_t(1,2) < 0.2$ GeV/c (d) $0.6 < p_t(\rho^o) < 1.5$ GeV/c, $p_t(1) > 0.2$ GeV/c, $p_t(2) < 0.2$ GeV/c (e) $1.5 < p_t(\rho^o) < 5.5$ GeV/c, $p_t(1) > 0.2$ GeV/c, $p_t(2) < 0.2$ GeV/c.; (e) $5.5 < p_t(\rho^o) < 10.$ GeV/c, $p_t(1,2) < 2.0$ GeV/c. In Figure 4 (a), the $\rho^0$ are essentially produced at rest in the laboratory frame, one then observes that the correlation is narrowly peaked at 180º. In Figure 4 (b-f), one progressively increases the momentum of the decaying $\rho^0$ thereby resulting in a kinematical focusing of the pions produced by the decays. One thus finds that the correlation function progressively broadens, and develops a dip at 180º. When the $\rho^0$ are decayed at "high" momentum, the angle of separation between the pions becomes small and leads to a narrow correlation function peaked at 0º as shown in Figure 4 (e-f).

It is obvious from Figure 4 that even a simple phenomenon such a rho-meson decay can produce a wide variety of correlation function shapes which are determined in part by the kinematics of the decay, and in part by the dynamics of the decaying particle (in our example the momentum of the $\rho^0$). While this example may seem trivial, we stress that the production of particles in di-jet events may behave similarly as in Figure 4. Indeed, consider that "flowing di-jets" could lead to similar kinematical focusing and consequently, correlation functions that depend on the di-jet velocity in the laboratory frame.

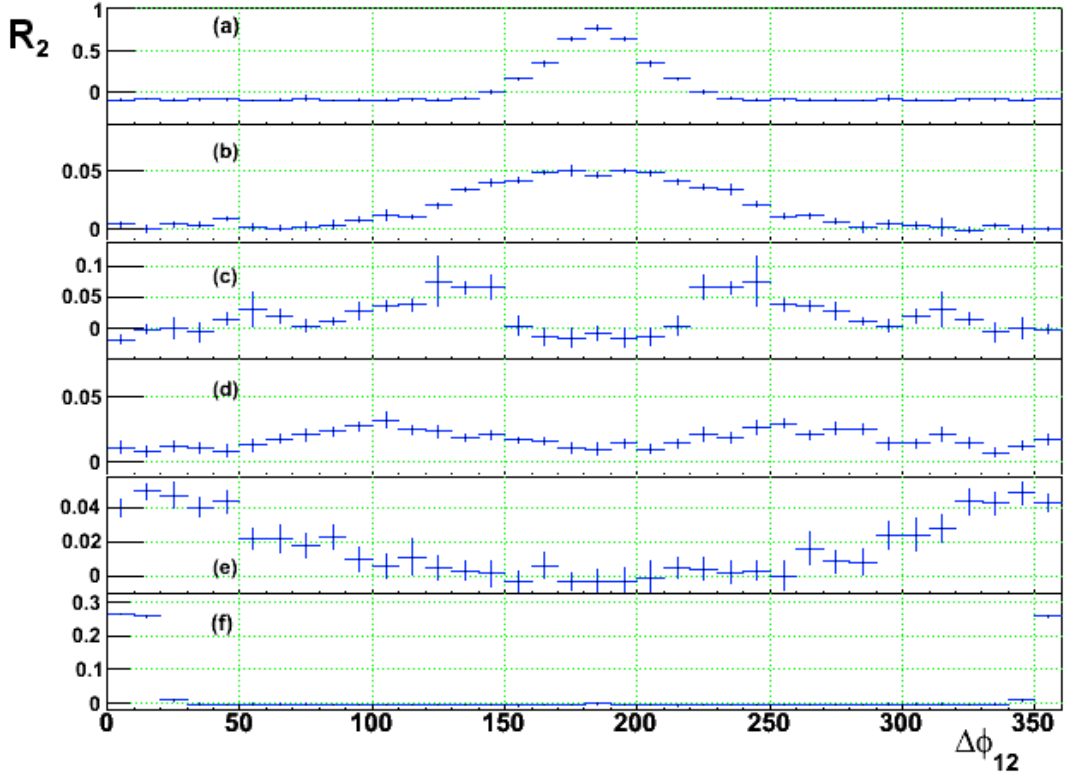

**Figure 4** Two-particle correlations obtained with a toy model simulating the decay of $\rho^0$ mesons in selected momentum ranges (a) $0.01 < p_t(\rho^0) < 0.1$ GeV/c, $p_t(1,2) < 0.2$ GeV/c; (b) $0.1 < p_t(\rho^0) < 0.5$ GeV/c, $p_t(1) > 0.3$ GeV/c, $p_t(2) < 0.2$ GeV/c (c) $0.1 < p_t(\rho^0) < 0.5$ GeV/c, $p_t(1,2) < 0.2$ GeV/c (d) $0.6 < p_t(\rho^0) < 1.5$ GeV/c, $p_t(1) > 0.2$ GeV/c, $p_t(2) < 0.2$ GeV/c (e) $1.5 < p_t(\rho^0) < 5.5$ GeV/c, $p_t(1) > 0.2$ GeV/c, $p_t(2) < 0.2$ GeV/c.; (e) $5.5 < p_t(\rho^0) < 10.$ GeV/c, $p_t(1,2) < 2.0$ GeV/c. (in color online)

We conclude this section with an example of three-particle cumulant applied to two-body decays. We use this example to illustrate the power of the cumulant technique, and to show its application in a practical case. Obviously, for two-body decays, no signal should be found in the 3-cumulant. We explicitly demonstrate this point with a simple simulation. Our example is based on an arbitrary (and unphysical) mix of $\rho^0$, $\pi^+$ and $\pi^-$ in relative abundance of 1:0.5:0.5. Primary pions are produced with a thermal (T=0.4 GeV) spectrum. $\rho(770)$ are produced at $90°$ from the beam direction, with a transverse momentum in the range $0.1 < p_t < 0.5$ GeV/c, and decayed into pairs of $\pi^+$ and $\pi^-$. Events are produced with random multiplicity ranging from 30 to 50 particles per event. Figure 5 (a) displays the normalized three-particle density obtained with a sample of four million events while requiring particle 1 to have a pt greater than 0.3 GeV/c, and particles 2 and 3 to have a pt smaller than 0.3 GeV/c. Although the physical phenomenon involved is relatively simple (two pions emitted back-to-back in the lab frame) the three-particle density is rather complicated. The apparent complexity stems from the fact the three-particle density is a superposition of many terms. With the chosen $\rho^0$ momentum range, and the pt cuts used in the production of the plot, one gets correlated particles from combining particles 1 & 2, 1 & 3, or 2 & 3. This is explicitly shown in Figures 5 (b-d) which display the normalized combinatorial terms $\rho_2\rho_1(12,3)/\rho_1\rho_1\rho_1$, $\rho_2\rho_1(13,2)/\rho_1\rho_1\rho_1$, $\rho_2\rho_1(23,1)/\rho_1\rho_1\rho_1$. The terms $\rho_2\rho_1(12,3)/\rho_1\rho_1\rho_1$ and $\rho_2\rho_1(13,2)/\rho_1\rho_1\rho_1$ exhibit a strong back-to-back correlation, i.e., a peak at $180°$, between particles 1&2, and 1&3 respectively. $\rho_2\rho_1(23,1)/\rho_1\rho_1\rho_1$ shows a weaker, yet finite back-to-back correlation. Note these three plots account for all correlations shown in Fig 5 (a). Indeed, one finds the three-particle cumulant, shown in Fig 5 (e) is flat and featureless, thereby indicating there are no three-particle correlation signals in the analyzed data. Interestingly, while the three-particle density has an apparently complicated structure, the actual correlation is, in fact, a null signal (as it should be).

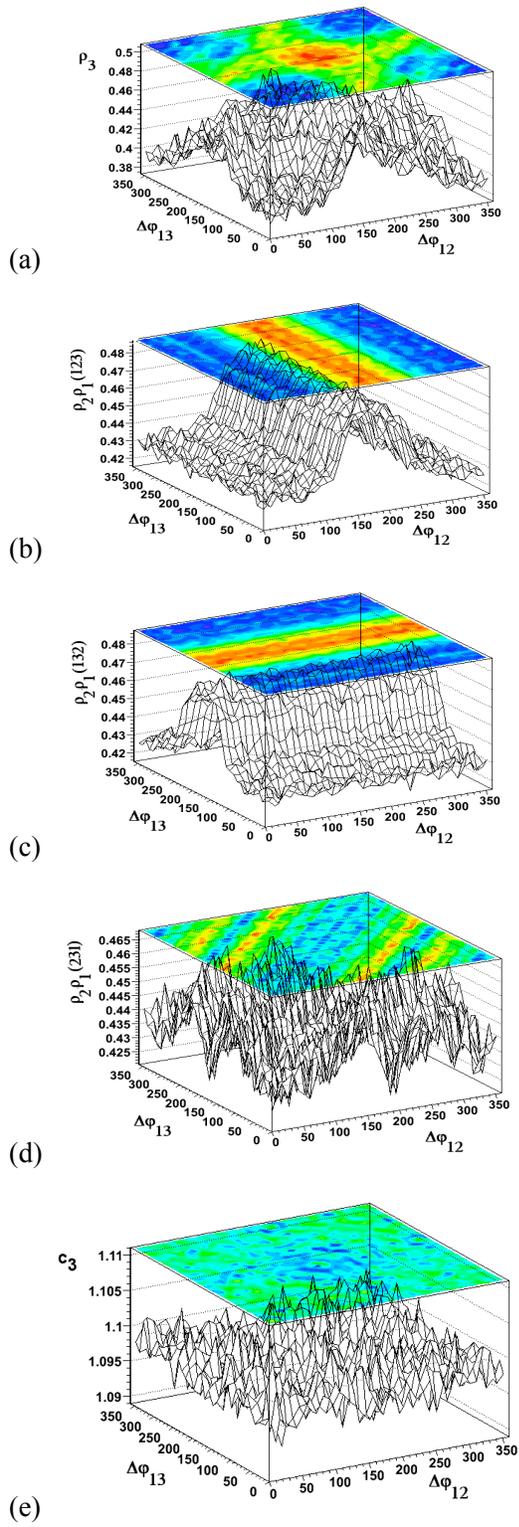

**Figure 5 (a) Normalized three-particle density, $\rho_3(\varphi_{12}, \varphi_{13})/\rho_1\rho_1\rho_1$ obtained for events containing a mixture of $\rho^o$, $\pi^+$, and $\pi^-$; (b-d) combinatorial terms $\rho_2\rho_1(12,3)/\rho_1\rho_1\rho_1$, $\rho_2\rho_1(13,2)/\rho_1\rho_1\rho_1$, $\rho_2\rho_1(23,1)/\rho_1\rho_1\rho_1$; (e) cumulant $C_3(\varphi_{12}, \varphi_{13})$.**

### 3.7. Jet Differential Attenuation (JDA)

We explore the possibility jet production in A + A non-central collisions may be modulated in azimuth relative to the reaction plane. This modulation may arise, for instance, from finite jet parton attenuation length in the produced medium, as schematically illustrated in Figure 6. The penetration of the parton through the medium is subject to differential attenuation (which depends on the medium density) and the parton interaction cross-section. This implies the jet produced by the fragmentation of these partons may exhibit finite azimuthal dependency relative to the reaction plane. This simple scenario neglects the possible disturbance imparted to the medium by the propagation of the jet (or parton).

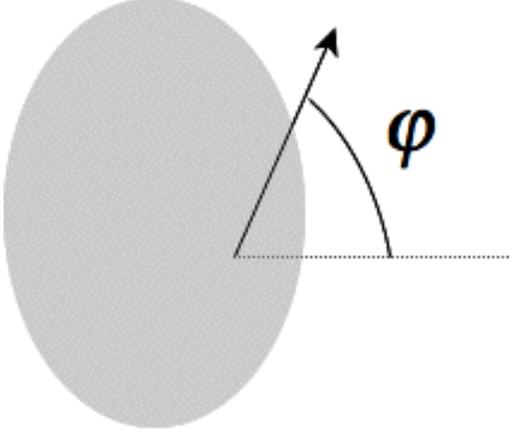

Figure 6 Schematic of Jet differential azimuthal attenuation through the dense medium produced in A+A collisions.

We model the possible jet dependency on azimuthal angle relative to the reaction plane with a Fourier series. Specifically, we write the probability of the jet being emitted at angle $\phi$ while the reaction plane is at $\psi$ as:

$$P(\phi,\psi) = 1 + 2\sum_m a_m \cos(m(\phi-\psi)) \tag{3.39}$$

where the coefficients represent the effect of the differential azimuthal attenuation.

The dependence on emission angle relative to reaction plane is known as event anisotropy or flow. Flow arises in this context from differential attenuation of the initial parton, but it may also arise from pressure gradients. Whether these two sources can be disentangled is an open issue and perhaps a matter of definition. In order to keep this model relatively simple, we assume the jet hadronization occurs outside the medium or is not affected by its presence. We, therefore, can parameterize the jet multiplicity and azimuthal width as in Section 3.2 using associated yields, and Gaussian widths that do not depend on the azimuthal direction. Thus, only the number of jets is considered to vary with azimuthal angle relative to the reaction plane. The form of the single particle density is unchanged relative to that found in Sect. 3.2:

$$\rho_{1,MJ}(\varphi_i) = (2\pi)^{-1} \langle J \rangle \langle A_i \rangle \tag{3.40}$$

The two-particle density is modified by the differential attenuation. One finds:

$$\rho_{2,MJ}(\varphi_i,\varphi_j) = (2\pi)^{-1} \langle J \rangle \langle A_i A_j \rangle P_{2,G}(\varphi_i,\varphi_j;\sigma_i,\sigma_i)$$
$$+ (2\pi)^{-1} \langle J(J-1) \rangle \langle A_i \rangle \langle A_j \rangle P_{2,MIX}(\varphi_i,\varphi_j;\sigma_i,\sigma_i) \tag{3.41}$$

where $P_{2G}$ is given by equation 3.13. The term $P_{2MIX}$ accounts for correlations between particles belonging to different jets caused by differential azimuthal attenuation.

$$P_{2,MIX}(\varphi_i,\varphi_j;\sigma_i,\sigma_j,a_m) = 1 + 2\sum_m a_m(i)a_m(j)\exp\left(-\frac{m^2}{2}(\sigma_i^2+\sigma_j^2)\right)\cos(m(\varphi_i-\varphi_j)) \quad (3.42)$$

The cos(m($\varphi_i$-$\varphi_j$)) dependency arises from two-jet convolution and constitutes a flow-like signal. It is important to realize that while this flow signal results from differential attenuation, it is in practice indistinguishable from flow produced by other mechanisms (e.g., pressure gradient), unless one also models its transverse momentum dependence. Note the strength of the modulation depends on both the anisotropy coefficients $a_m$ and the widths of the jets $\sigma_i$. Thus, one finds the harmonic coefficients $v_m$ of expression (3.42) are given by:

$$v_m(i) = a_m(i)\exp\left(-\frac{m^2}{2}\sigma_i^2\right) \quad (3.43)$$

The calculation of the three-particle density yields.

$$\rho_{3,MJ}(\varphi_i,\varphi_j) = (2\pi)^{-1}\langle J\rangle\langle A_i A_j A_k\rangle P_{3,G}(\varphi_i,\varphi_j,\varphi_k;\sigma_i,\sigma_j,\sigma_k)$$
$$+(2\pi)^{-1}\langle J(J-1)\rangle\left\{\begin{array}{l}\langle A_i A_j\rangle\langle A_k\rangle P_{2,1,MIX}(\varphi_i,\varphi_j;\sigma_i,\sigma_j)\\+\langle A_i A_k\rangle\langle A_j\rangle P_{2,1,MIX}(\varphi_i,\varphi_k;\sigma_i,\sigma_k)\\+\langle A_j A_k\rangle\langle A_j\rangle P_{2,1,MIX}(\varphi_j,\varphi_k;\sigma_j,\sigma_k)\end{array}\right\} \quad (3.44)$$
$$+(2\pi)^{-1}\langle J(J-1)\rangle\langle A_i\rangle\langle A_j\rangle\langle A_k\rangle P_{3,MIX}(\varphi_i,\varphi_j,\varphi_k;\sigma_i,\sigma_j,\sigma_k)$$

where $P_{3GJ}$, given by equation 3.13, accounts for the jet components, while the $P_{2,1,MIX}$ and $P_{3,MIX}$ functions account for correlation caused by the differential attenuation. One finds:

$$P_{2,1,MIX}(\varphi_i,\varphi_j;\sigma_i,\sigma_j,a_m) = P_{2,G}(\varphi_i,\varphi_j;\sigma_i,\sigma_j)$$
$$\times\left\{1+2\sqrt{2\pi}\sum_m a_m(i)a_m(j)\exp\left(-\frac{m^2\sigma_{ijk}^4}{2\sigma_{ij}^2}\right)\cos\left(m\left(\frac{\sigma_j^2\varphi_i+\sigma_i^2\varphi_j}{\sigma_i^2+\sigma_j^2}-\varphi_j\right)\right)\right\} \quad (3.45)$$

$$\sigma_{ij}^2 = \sigma_i^2 + \sigma_j^2$$
$$\sigma_{ijk}^4 = \sigma_i^2\sigma_j^2 + \sigma_i^2\sigma_k^2 + \sigma_j^2\sigma_k^2$$

The $P_{3MIX}$ function reduces to :

$$P_{3,MIX}(\varphi_i,\varphi_j,\varphi_k;\sigma_i,\sigma_j,\sigma_k,a_m) = 1$$
$$+2\sum_m v_m(i)v_m(j)\cos(m(\varphi_i-\varphi_j))$$
$$+2\sum_m v_m(i)v_m(k)\cos(m(\varphi_i-\varphi_j)) \quad (3.46)$$
$$+2\sum_m v_m(j)v_m(k)\cos(m(\varphi_i-\varphi_j))$$
$$+2\sum_{m,n,p} v_m(i)v_m(j)v_m(k)\times\left[\begin{array}{l}\delta_{m,n+p}\cos(m\varphi_i+n\varphi_j+p\varphi_k)\\+\delta_{n,m+p}\cos(m\varphi_i+n\varphi_j+p\varphi_k)\\+\delta_{p,n+m}\cos(m\varphi_i+n\varphi_j+p\varphi_k)\end{array}\right]$$

where $v_m$ coefficients are given by expression (3.43).
The last term of $P_{3mix}$ is identical in form to the non-diagonal irreducible flow terms found in expression (3.25). Thus, one concludes the differential attenuation produces a flow-like signal even in the three-particle density. The jet attenuation, however, also produces cross-harmonic components.

Indeed, one finds the P$_{2,1\text{MIX}}$ term contains a dependence on $\cos\left(m\left(\frac{\sigma_j^2\varphi_i + \sigma_i^2\varphi_j}{\sigma_i^2 + \sigma_j^2} - \varphi_k\right)\right)$ which for $\sigma_i^2 = \sigma_j^2$ can be written $\cos\left(\frac{m}{2}(\Delta\varphi_{ik} + \Delta\varphi_{jk})\right)$. However, this cosine dependence enters as a coefficient of P$_{2G}$ and, as such only causes a modification of the jet shape (here arbitrarily assumed to be Gaussian) that could be difficult to observe in practice.

The three-particle cumulant is obtained from above equations.

$$\hat{\rho}_{3,MJ}(\varphi_i,\varphi_j) = (2\pi)^{-1}\langle J\rangle\langle A_i A_j A_k\rangle P_{3,G}(\varphi_i,\varphi_j,\varphi_k;\sigma_i,\sigma_j,\sigma_k)$$

$$+ (2\pi)^{-1}\langle A_i A_j\rangle\langle A_k\rangle P_{GJ}\left(\langle J(J-1)\rangle - \langle J\rangle^2 + 2\langle J(J-1)\rangle\sum_m v_m(i)v_m(j)\cos(m(\varphi_i-\varphi_j))\right)$$

$+(i,j,k)$ permutations of above

$$+ (2\pi)^{-1}\langle A_i\rangle\langle A_j\rangle\langle A_k\rangle\left\{\langle J(J-1)(J-2)\rangle - 3\langle J(J-1)\rangle\langle J\rangle + 2\langle J\rangle^3\right\} \quad (3.47)$$

$$+ 2(2\pi)^{-1}\langle A_i\rangle\langle A_j\rangle\langle A_k\rangle\left\{\langle J(J-1)(J-2)\rangle - 3\langle J(J-1)\rangle\langle J\rangle\right\}\sum_m v_m(i)v_m(j)\cos(m(\varphi_i-\varphi_j))$$

$+(i,j,k)$ permutations of above

$$+ 2(2\pi)^{-1}\langle A_i\rangle\langle A_j\rangle\langle A_k\rangle\langle J(J-1)(J-2)\rangle \sum_{m,n,p} v_m(i)v_m(j)v_m(k) \times \begin{bmatrix} \delta_{m,n+p}\cos(m\varphi_i + n\varphi_j + p\varphi_k) \\ +\delta_{n,m+p}\cos(m\varphi_i + n\varphi_j + p\varphi_k) \\ +\delta_{p,n+m}\cos(m\varphi_i + n\varphi_j + p\varphi_k) \end{bmatrix}$$

The structure of this cumulant is rather complicated owing to the correlations induced by the presence of multiple jets within each event. We note that the differential attenuation produces non-reducible terms of order $v_2(i)v_2(j)$ as well as non-diagonal terms such as $v_2v_2v_4$. While the amplitude of these coefficients depend on the number of jets and associated fragments, we find their multiplicity dependence is essentially indiscernible from that obtained in Eq. (3.25) where we assumed the jet production is completely decoupled from the flowing bulk. Note that the above contains a modulation of the jet shape (2$^{nd}$ line) not found in Eq. (3.22) but this effect is likely to be difficult to observe in practice because the Gaussian jet approximation used in this simple model is not necessarily justified.

### 3.8. Mach Cone Emission

Mach cone emission of particles by partons propagating through dense QGP matter was proposed by Shuryak[25] to explain the peculiar dip structure found at 180° in two-particle correlations recently reported by the PHENIX collaboration. Mach cone emission was also discussed by Stoecker[26], and more recently by a number of other authors[14,27]. Cerenkov gluon emission has been proposed as an alternative explanation of the PHENIX data[28]. Essentially, the concept of Mach cone emission is based on the notion that high momentum partons propagating through a dense QGP interact with the medium and lose energy (and momentum) at a finite rate. The release of energy engenders a wake that propagates at a characteristic angle (the Mach angle) determined by the sound velocity in the medium. Shuryak[25] estimates the speed of sound in the QGP to be on the order of $v_s \sim \bar{c}_s^{RHIC} \approx 0.33$. The Mach angle is thus expected to be on the order of 70° relative to the away-side parton direction.

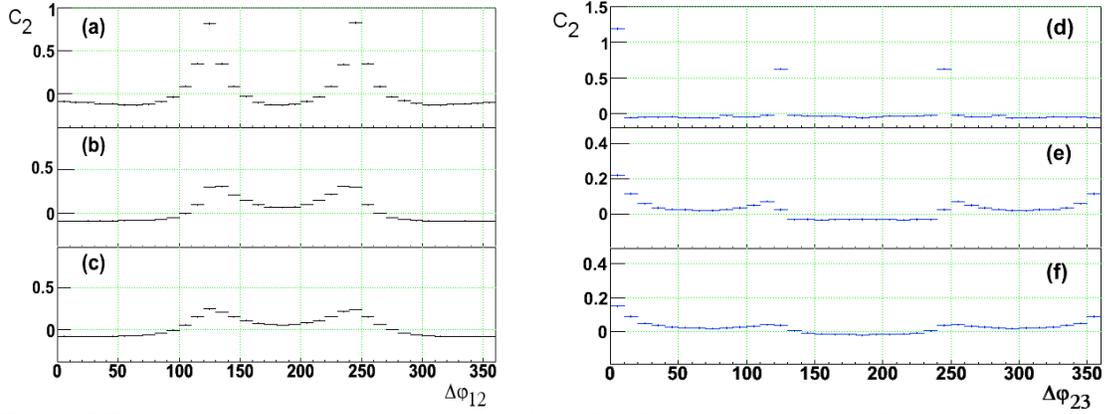

Figure 7 Two-particle cumulants obtained with the Mach cone models described in the text. Plots (a-c) show correlations between the trigger particle and an associate (low pt) particle emitted in the Mach cone. Plots (d-f) show correlations between associate (Mach cone) particles emitted with a low pt. Plots (a) and (d) correspond to Mach cone particle emission strictly perpendicular to the beam direction. Plots (b) and (e) are for full Mach cone emission (all azimuths relative to the away-side direction) for an away-side parton emitted at $90^o$ from the beam direction. Plots (c) and (f) are for uniform away-side emission (color online).

While the concept of Mach cone is simple, its realization in two- or three-particle correlations is perhaps not as intuitive as one might think. We illustrate this point with three simple geometric models of increasing realism. In all three models, the near-side jet is reduced, for simplicity, to one particle (hereafter called trigger) emitted with a fixed transverse momentum of 3 GeV/c, with a Gaussian profile of $10^o$ width; While the away-side particles, produced with fixed $p_t$ of 1 GeV/c, are assumed to consist of Mach cone particles only. We first present, in Figures 7 (a, d), two-particle cumulants obtained with trigger and Mach cone particles emitted at $90^o$ from the beam direction. Three-particle cumulants are shown in Figure 8. For the construction of these correlations, we use a requirement of $p_t > 2$ for particle 1, and $p_t < 2$ for particles 2 and 3 (i.e., particles 2 and 3 are exclusively from the Mach cone). Given that Mach cone particles are produced at $\pm 60^o$ from the away-side direction and strictly normal to the beam direction, two and four narrow peaks are respectively seen in the two- and three-particle correlations. One finds that, as suggested by many authors, a strong dip is present at $180^o$ in the two-particle correlations, while in the three-particle correlation a clear spacing is found between the peaks. Note that the finite width of the peaks is due, in this simple model, to the finite width of the trigger jet. In practice, one might expect additional broadening of the cone because the speed of sound changes through the life of the QGP medium, and given the finite size of the medium. It is also highly unlikely cone emission should be restricted to directions perpendicular to the beam-axis. We thus relax this requirement and present correlations obtained when Mach cone particles are emitted at $60^0$ from the away-side direction including all azimuths in Figures 7 (b,e) and in Figure 8 (b). Here one finds the depth of the dip is dramatically reduced in the two-particle correlation, while the space between the four peaks of the three-particle correlation is now partially filled. This second model, however, is rather unrealistic. Jet emission is not restricted to normal angles relative to the beam direction and proceeds in a large range of rapidities. We include emission over an extended range of rapidities, with uniform probability trigger distribution in the range of $|\cos\theta|<0.7$, in our next simulation results, shown in Figure 7 (c,f) and Figure 8 (b). One observes the projection of the cone on the transverse plane leads to a slight broadening of the peaks structures found in the previous model. One concludes the details of the correlation shapes clearly depend on assumptions made about the kinematics of the away-side parton. Also note that we have assumed in these three simple models that the Mach cone emission occurs at a very specific angle (i.e. $60^o$). In practice, one should expect both the sound velocity is a function of the local density of the medium, as well as of the parton velocity. The cone particles may also themselves be deflected by the bulk, thereby leading to additional smearing of the peak structures.

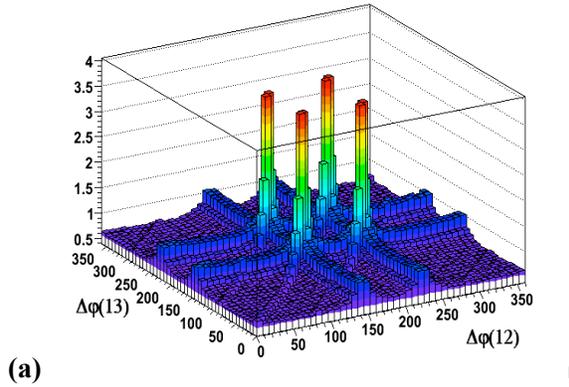

(a)

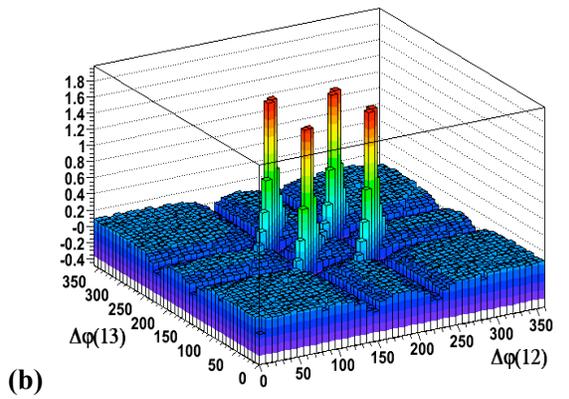

(b)

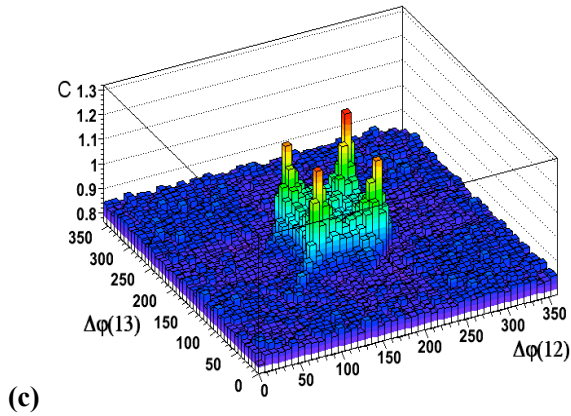

(c)

**Figure 8** (a) Normalized three-particle density obtained with the Mach cone model described in the text; (b) normalized three-particle cumulant for Mach cone emission in the transverse plane only; (c) normalized three-particle cumulant for Mach cone emission in all azimuths relative to the away-side direction. (Color online)

## Discussion and Summary

We presented a new technique based on cumulants for the analysis of three-particle distributions designed to distinguish between different particle production mechanisms. We argued that while two-particle correlations have enabled the identification of interaction of jets with the medium in Au + Au collisions at RHIC, ambiguities are left in the interpretation of some of the correlation functions reported by PHENIX and STAR. Indeed, it is possible to explain the broadening and dip observed in away-side jet structures on the basis of away-side deflection, di-jet flow, as well as Mach cone emission. Using specific toy models, we showed these processes lead to distinguishable features in the analysis of three-particle azimuthal correlations based on the cumulant technique. Our discussion is based on three-particle correlations plotted as a function of two azimuthal angle differences $\Delta\varphi_{12} = \varphi_1 - \varphi_2$ and $\Delta\varphi_{13} = \varphi_1 - \varphi_3$ where $\varphi_1$, $\varphi_2$, and $\varphi_3$ are the azimuthal angle of emission of particles considered in the analysis. We showed that if one chooses particle 1 to be a high $p_t$ particle, and particles 2 &3 lower $p_t$ particles, one effectively becomes sensitive to a leading jet particle and particles associated with the same or away-side jet. We showed that scattering of the away-side parton results in a broadening of the away-side jet correlation peak along the main diagonal of the $\Delta\varphi_{12}$ and $\Delta\varphi_{13}$ correlation plane while away-side jet broadening due to interactions with the medium produces a broadening of the away-side jet correlation peak along the second diagonal of the $\Delta\varphi_{12}$ and $\Delta\varphi_{13}$ correlation plane (proportional to $\Delta\varphi_{23}$) as well as the main diagonal. By contrast, Mach cone or Cerenkov emission should lead to four peak structures in the $\Delta\varphi_{12}$ and $\Delta\varphi_{13}$ correlation plane: two along the main diagonal, and two along the second diagonal (provided the transverse momentum ranges used to select particles 1 and 2&3 are suitably selected to identify hard particles from a jet, and "soft" particles from Mach cone emission). We remark, however, that if the $p_t$ range of particle 1 is lowered to include Mach cone emission, additional structures shall appear in the correlation function.

We discussed in detail how the presence of anisotropic flow influences the three-particle density. We derived expressions for harmonic flow terms in the two- and three-particle cumulants and showed second order terms in $v_2 v_2$ are naturally removed from the three-particle cumulants, while non-diagonal, higher order "irreducible" terms persist. We argued that non-diagonal terms should be dominated by $v_2 v_2 v_4$ terms. Such non-diagonal terms can, however, be modeled and explicitly subtracted based on measured values of $v_2$ and $v_4$. The three-particle cumulant technique presented in this paper enables straightforward and unambiguous elimination of flow effects and thereby a robust measurement of jet-like features. Thus, in spite of the fact that low $p_t$ jets cannot be identified on an event-by-event basis in heavy ion collisions, it is possible to gain detailed insight of jet interactions with the medium produced in these collisions.

## Acknowledgements

This work was supported, in part, by U.S. DOE Grant No. DE-FG02-92ER40713. The author acknowledges fruitful discussions with Drs. S. Gavin, I. Selyuzhenkov, P. Stankus, and S. Voloshin. The author also thanks J. Ulery and F. Wang for providing a $v_2$ parameterization of STAR data on elliptical flow.